%% file: main.tex
\documentclass[11pt]{article}

\usepackage{fullpage}
\usepackage[ruled, linesnumbered, vlined, commentsnumbered]{algorithm2e}
\usepackage{graphicx,subcaption} 
\usepackage{amsfonts,amssymb,amsmath,amsthm,amsopn}	
\usepackage{hhline,booktabs,colortbl,multirow,tabularx,diagbox,threeparttable} 
\usepackage[listings,skins,breakable]{tcolorbox}

\usepackage{enumerate}
\usepackage{authblk}
\usepackage{footnote}
\usepackage{hyperref}
\usepackage{prettyref}
\usepackage{cite}
\usepackage{setspace}
\usepackage{color}
\usepackage{xcolor}  
\usepackage{lscape}
\usepackage{geometry}
\usepackage{relsize}

\usepackage{tikz}
\usepackage{pgfplots}
\usetikzlibrary{positioning,shapes,shadows,arrows,calc}
\tikzstyle{component}=[rectangle, draw=black, rounded corners, fill=blue!40, drop shadow, text centered, anchor=north, text=white, minimum height=1cm]
\tikzstyle{arrow}=[->, thick]

\pgfplotsset{compat=1.12}
\usetikzlibrary{intersections}
\usetikzlibrary{pgfplots.statistics}
\usepgfplotslibrary{fillbetween}

\tcbset{colback=white, arc=0mm, outer arc=0pt}

\def\hlinew#1{%
  \noalign{\ifnum0=`}\fi\hrule \@height #1 \futurelet
   \reserved@a\@xhline}
\makeatother

\definecolor{myblue}{RGB}{34,31,217}
\definecolor{mycyan}{gray}{.7}
\definecolor{Gray}{gray}{0.9}

\newtheorem{remark}{Remark}

\newtheorem{lemma}{Lemma}

\let\oldnl\nl
\newcommand{\noln}{\renewcommand{\nl}{\let\nl\oldnl}}

\newcommand{\pref}{\prettyref}

\newrefformat{fig}{Fig.~\ref{#1}}
\newrefformat{tab}{Table~\ref{#1}}
\newrefformat{sec}{Section~\ref{#1}}
\newrefformat{app}{Appendix~\ref{#1}}
\newrefformat{alg}{Algorithm~\ref{#1}}
\newrefformat{property}{Property~\ref{#1}}
\newrefformat{theorem}{Theorem~\ref{#1}}
\newrefformat{corollary}{Corollary~\ref{#1}}
\newrefformat{proposition}{Proposition~\ref{#1}}
\newrefformat{def}{Definition~\ref{#1}}
\newrefformat{eq}{equation~(\ref{#1})}

\newcolumntype{R}{>{\raggedleft\arraybackslash}X}
\newcolumntype{L}{>{\raggedright\arraybackslash}X}
\newcolumntype{P}[1]{>{\raggedright\arraybackslash}p{#1}}

\newcommand{\abs}[1]{\left\lvert #1 \right\rvert}

\makeatletter
\def\BState{\State\hskip-\ALG@thistlm}
\makeatother

\SetCommentSty{mycommfont}

\title{\vspace{-1ex}\LARGE\textbf{Evolutionary Multi-Objective Optimization for Virtual Network Function Placement}\footnote{This manuscript is submitted for potential publication. Reviewers can use this version in peer review.}}

\author[1]{\normalsize Joseph Billingsley}
\author[1]{\normalsize Ke Li}
\author[1]{\normalsize Geyong Min}
\author[2]{\normalsize Nektarios Georgalas}
\affil[1]{\normalsize Department of Computer Science, University of Exeter, EX4 4QF, Exeter, UK}
\affil[2]{\normalsize Research and Innovation, British Telecom, Martlesham, UK}
\affil[$\ast$]{\normalsize Email: \texttt{k.li@exeter.ac.uk}}

\date{}

\begin{document}
\maketitle

\vspace{-3ex}
{\normalsize\textbf{Abstract: }Data centers are critical to the commercial and social activities of modern society but are also major electricity consumers. To minimize their environmental impact, it is imperative to make data centers more energy efficient while maintaining a high quality of service (QoS). Bearing this consideration in mind, we develop an analytical model using queueing theory for evaluating the QoS of a data center. Furthermore, based on this model, we develop a domain-specific evolutionary optimization framework featuring a tailored solution representation and a constraint-aware initialization operator for finding the optimal placement of virtual network functions in a data center that optimizes multiple conflicting objectives with regard to energy consumption and QoS. In particular, our framework is applicable to any existing evolutionary multi-objective optimization algorithm in a plug-in manner. Extensive experiments validate the efficiency and accuracy of our QoS model as well as the effectiveness of our tailored algorithms for virtual network function placement problems at various scales.


\input{introduction}

\input{related}

\input{problem_formulation}

\input{system_model}

\input{optimisation}

\input{experiments}

\input{conclusions}

\section*{Acknowledgment}
K. Li was supported by UKRI Future Leaders Fellowship (MR/S017062/1, MR/X011135/1), EPSRC (2404317), NSFC (62076056), Royal Society (IES/R2/212077) and Amazon Research Award.

\bibliographystyle{IEEEtran}
\bibliography{IEEEabrv,bibliography}

\end{document}

%% file: introduction.tex
\section{Introduction}
\label{sec:introduction}

Recent research indicate that data centers will be responsible for 3\% to 5\% of total energy consumption worldwide by 2030~\cite{AndraeE15}. With the pressing need to address climate change, there are environmental as well as business imperatives to improve the efficiency of data centers wherever possible. Over the past decade, data centers have become significantly more energy efficient by reducing overhead such as heat management and energy provisioning~\cite{AvgerinouBC17}. Despite these efforts, the total energy consumed by data centers were still doubled between 2010 and 2020 due to an increased demand, and there are diminishing returns to reducing overhead further~\cite{DoddAGC20}. A recent study showed that future efficiency improvements can be made by using fewer network components and better operational policies~\cite{DoddAGC20}. One route to achieve this is through virtualization, i.e., the emulation of hardware with software. Physical computing devices can be virtualized into virtual machines (VMs), and several VMs can be executed on a single physical device. By placing applications on VMs and packing multiple VMs onto the same server, we can maximize the utilization of hardware and consume less energy to provide the same quality of service (QoS). In addition, VMs can be moved and scaled to meet traffic demands without over or under allocating resources. A recent study found that simply utilizing servers more effectively with virtualization would result in a 10\% reduction in data center energy consumption in the USA~\cite{ShehabiARSSD16}. This reduction increases to 40\% if the majority of service providers move to \lq hyper-scaled\rq\ data centers which have more powerful servers with a larger capacity.

Virtualization was applied to general purpose servers that contribute some of the computing power required to provide services. More recently, purpose-built network functions have also been considered as targets for virtualization. A network function is a network component that performs a specific task such as load balancing or packet inspection. Services, such as phone call handling or video streaming, usually direct traffic through several network functions in a prescribed order. Traditionally, these functions were provided by \lq middleboxes\rq\ through purpose-built hardware. However, middleboxes cannot be scaled or moved like VMs thus limiting the flexibility of the data center. Virtual network functions (VNFs) provide the same functionality as middleboxes but with software running on VMs. Although each VNF instance may perform relatively worse than its equivalent middlebox, the added flexibility can improve the overall performance and reduce costs.

In a nutshell, a VNF placement problem (VNFPP) aims to find the optimal number and placement of VNFs in order to optimize the QoS (e.g., minimizing the expected latency and packet loss) of each service, balanced against the energy consumption of the data center. A VNFPP instance defines a set of services and a data center topology. Each service is defined by its packet arrival rate and a service chain, i.e., the sequence in which VNFs must be visited. A solution to the VNFPP defines where to place VNFs for each service and how packets should traverse the data center. The VNFPP has been widely recognized as a challenging combinatorial optimization problem given its NP-hardness~\cite{LuizelliCBG17,SangJGDY17,CohenLNR15}, multiple conflicting objectives and a proportionally small feasible solution space. Although many efforts have been devoted to this problem, there are three unsolved challenges ahead.
\begin{itemize}
    \item The first one lies in the QoS evaluation itself. There exist some tools, such as discrete event simulators that can provide accurate QoS measurements. However, they are too time consuming to be incorporated into an optimization routine. In contrast, some heuristics, such as the number of applied VNF instances~\cite{LuizelliCBG17,AddisBBS15} and the average utilization of servers~\cite{JemaaPP16,GaoABS18}, have been proposed as an efficient surrogate of the QoS. Yet, there is no established evidence that supports the equivalence of using such heuristics versus an accurate QoS measurement. In addition, there have been some attempts of using queuing theory, which has been widely recognized as a powerful tool to produce fast and accurate QoS models~\cite{LakshmiI2013,PapadopoulosC96}, for VNFPP~\cite{OljiraGTB17,MarottaZDK17,LeivadeasFLIK18,BillingsleyLMMG19}, packet loss and its consequences have been ignored, thus limiting their accuracy.   
    \item Second, the curse-of-dimensionality has been the Achilles' heel of optimization methods. For example, linear programming (LP), one of the most popular methods for VNFPP, is only useful for problems with tens or hundreds of servers~\cite{BariCAB15,KawashimaOOM16,AllegKMA17}. Meta-heuristics have recently shown some encouraging results on a larger-scale VNFPP with up to $1,000$ servers~\cite{LuizelliCBG17}. However, none of them are close to industrial-scale scenarios.
    \item Last but not the least, a VNFPP usually has complex constraints, such as the routing constraints that require the solution to visit VNFs in a prescribed order and a limit of the number of instances of each VNF or where they can be placed. These constraints significantly squeeze the feasible search space thus hindering an effective search.
\end{itemize}

Evolutionary algorithms (EAs) have been well recognized to be effective for tackling multi-objective optimization problems (MOPs) \cite{WangJ20,ZhouL17,WangWZ19,LiuLJ14,LiDZK15,LiCFY19,LiKZD15,LiZKLW14,LiFKZ14}. However, few works have considered their applications to the VNFPP~\cite{CaoZACHS16,RankothgeLRL17,LangeGZTJ17,BillingsleyLMMG19,BillingsleyLMMG20,BillingsleyMLMG20,BillingsleyLMMG21}. In this work, we provide a domain-specific evolutionary optimization framework to address the above longstanding problems. Our major contributions are as follows.
\begin{itemize}
    \item By using a queueing theory, we developed an analytical model that provides an efficient and accurate way to evaluate the QoS with regard to the expected latency, the packet loss of each service and the overall energy consumption of the underlying data center, all of which constitute the three-objective VNFPP in this paper.
    \item We developed a problem-specific solution representation for the VNFPP along with a tailored initialization operator that together promote a fast convergence and a feasibility guarantee. Both operations can be seamlessly incorporated into any evolutionary multi-objective optimization (EMO) algorithm.
    \item We validate the effectiveness and accuracy of the proposed algorithm under various settings. In particular, we consider problems with up to $8,192$ servers, which is $8\times$ larger than all reported results. The performance of our tailored EMO algorithms are compared to their generic counterparts as well as state-of-the-art (SOTA) heuristics.
\end{itemize}

In the remainder of this paper, \pref{sec:lit_review} provides a pragmatic overview of some selected developments on VNFPP. \pref{sec:problem_formulation} gives our VNFPP definition followed by a rigorous derivation of our analytical model in~\pref{sec:system_model}. \pref{sec:optimisation} develops tailored evolutionary optimization framework for multi-objective VNFPP. The effectiveness of our proposed analytical model along with the tailored evolutionary optimization framework are validated in~\pref{sec:experiments}. Finally, \pref{sec:conclusion} concludes this paper and sheds some lights on future directions.

%% file: related.tex

\section{Related Works}
\label{sec:lit_review}

This section overviews some selected developments in VNFPP according to the type of its solver including \textit{exact}, \textit{heuristic} and \textit{meta-heuristic} methods, respectively.

\subsection{Exact Methods}
\label{sec:exact}

Exact methods are designed to produce solutions with theoretical optimality guarantees. They have an exponential worst-case time complexity~\cite{Landa-Silva13}, thus are usually limited to small-scale VNFPPs. Furthermore, exact methods typically require linear objective functions which contradicts the nonlinear nature of QoS. To resolve this issue, some researchers use a simplified latency model where the waiting time at a switch is constant whereas in practice the waiting time depends on the switch's utilization \cite{IntelDPDK,IntelPPP,OljiraGTB17}. Bari et al.~\cite{BariCAB15} proposed to use dynamic programming to minimize a linear model of the operational cost under a latency constraint. Likewise, \cite{MiottoLCG19}, Miotto et al. developed a NFV optimization framework that applies LP to minimize the number of VNF instances and the length of routes under a latency constraint.

An alternative option is to use piece-wise linearization to linearize accurate models of QoS. In an early work on VNFPP, Baumgartner et al.~\cite{BaumgartnerRB15} proposed to minimize the total cost of bandwidth and VNF placement while meeting latency constraints for each service. After performing piece-wise linearization, they applied LP to this problem. Oljira et al.~\cite{OljiraGTB17} used the same technique as in~\cite{BaumgartnerRB15} for modeling and optimization and additionally considered the virtualization overheads when calculating the latency at each VNF. In~\cite{AddisBBS15}, Addis et al. proposed two different models for VNFPP. One models the waiting time as a convex piece-wise linear function of the sum of arrival rates while the other sets the latency as a constant when it is below a threshold. Later, Gao et al.~\cite{GaoABS18} extended this work and proposed additional constraints for affinity and anti-affinity rules that require solutions to place certain VNFs on the same server or apart respectively. In~\cite{JemaaPP16}, Jemaa et al. proposed a VNFPP formulation where VNFs can only be placed either in a resource constrained cloudlet data center near the user or an unconstrained cloud data center. They use exact methods to optimize latency, cloudlet and cloud utilization simultaneously.

\subsection{Heuristic Methods}
\label{sec:heuristics}

In contrast to exact methods, heuristic methods attempt to find approximate solutions and usually use a surrogate models as an alternative measure of QoS. A common model is the use of the available link or server capacity as a proxy for the latency and energy consumption. Guo et al.~\cite{GuoWLQA0Y20} formulated a VNFPP that aims to minimize the link and server capacity of a solution and allow VNFs to be shared across services. They first pre-processed the network topology to find the most influential nodes according to the Katz centrality \cite{Katz53}. Then, VNFs are placed according to a Markov decision process with lower costs for reusing VNFs. To promote VNF reuse, only shareable VNFs can be placed on the most influential nodes. Likewise, Qi et al.~\cite{QiSW19} formulated a similar problem where the total links and server usage must be minimized. They used a greedy search to exploit the neighborhood of each server. Based on the same assumption, Qu et al.~\cite{QuASK17} proposed to place VNFs on the shortest path between the starting and ending servers. If the path cannot accommodate all VNFs, servers close to the path will then be considered.

Another model assumes the a constant waiting time when a packet visits a component and the relevant algorithms are designed to minimize the network latency, i.e., the total waiting time incurred for a packet when traversing VNFs. For example, Hawilo et al.~\cite{HawiloJS19} proposed a heuristic that places the most commonly used VNFs on the central nodes determined by the betweeness centrality. This increases the likelihood that a short route can be constructed for each service. In~\cite{VizarretaCMMK17}, Vizarreta et al. proposed to set the waiting time as a constant while keeping the starting and ending nodes fixed. They first find the route that has the lowest cost and satisfies the latency and robustness constraints. Then, the route is adjusted until it can accommodate each VNF of the service. Beck et al. \cite{BeckB15} used a similar model to optimize the average path length and bandwidth usage. The heuristic searches the servers up to a small number of hops away and places the next VNF of each service on the nearest server that can accommodate it. If no such server is available, the earlier VNFs are removed.

Some researchers proposed to first use heuristics to place VNFs and then use accurate models to evaluate the solutions. Although this provides additional information to the decision makers, it does not improve the quality of solutions. For example, Zhang et al. \cite{ZhangXLLGW17} proposed a best-fit decreasing method to place VNFs and used a simple queueing model to evaluate the solution. In~\cite{ChuaWZSH16}, Chua et al. proposed a heuristic that iterates over the servers and places each VNF of each service at the first server with a sufficient capacity. In order to evenly distribute traffic, the available capacity for each server is limited. If every server has been considered before placing all VNFs, the heuristic increases the available capacity and reiterates the servers. Gouareb et al.~\cite{GouarebFA18} proposed a three-part heuristic that first assigns VNFs with the greatest resource demands to the servers with the largest capacity. Then, it uses either horizontal or vertical scaling to satisfy demand before finding the shortest routes between VNFs to form services. The heuristic was found to produce solutions an order of magnitude worse than an exact solver that uses an accurate model.

There also exist some attempts that try to bridge the gap between heuristics and exact methods. For example, Marotta et al.~\cite{MarottaZDK17} proposed to combine heuristics and LP. They apply a heuristic to place VNFs and make these placements robust to changes in the required resources for each VNF. Thereafter, LP is applied to find routes between VNFs while ensuring the satisfaction of latency constraints for each service. However, since the network is not considered until the final step, it is not guaranteed to find a solution. Agarwal et al.~\cite{AgarwalMCD18} use LP to assign a confidence score for whether a VNF should be assigned to a server. Then they use a greedy heuristic that considers the confidence score and the available capacity of the server to find VNF placements.

\subsection{Meta-heuristic Methods}
\label{sec:meta-heuristics}

As a subset of heuristic methods, meta-heuristic methods have been widely used for NP-hard problems~\cite{XueZB13,MavrovouniotisM17,YuanBTZLL17,ChenZLGGZYCLZ19,YoonK13} including other real world problems with high numbers of variables \cite{JiaMZ21,PengJW19,ChengJ15}. Yet, few studies can be found for VNFPPs. In~\cite{RankothgeMLRL15}, Rankothge et al. proposed a genetic algorithm (GA) to optimize VNF placement and routing by minimizing the number of servers and switches. In~\cite{CaoZACHS16}, Cao et al. used GA to minimize the bandwidth consumption and maximize the link utilization with a binary matrix solution representation for VNF placement and routing decisions. In~\cite{ChantreF20} and~\cite{KaurGK020}, a similar binary string representation is applied in multi-objective GAs. Specifically, \cite{ChantreF20} applied NSGA-II \cite{DebAPM02} to place primary and backup VNFs in small data centers while~\cite{KaurGK020} explored the effectiveness of different multi-objective GAs on a variety of QoS indicators. In~\cite{LangeGZTJ17}, a Pareto simulated annealing method is applied to find a set of trade-off solutions that optimize several indicators including a linear model of the expected latency, the number of hops, the number of VNF instances and the CPU utilization. Soualah et al.~\cite{SoualahMGZ17} proposed to use a Monte Carlo tree search to place VNFs and find routes between them so as to minimize the expected server utilization.

To the best of our knowledge, our previous work~\cite{BillingsleyLMMG19,ChenLY18,ZouJYZZL19,LiZZL09,LiZLZL09,Li19,LiK14,LiFK11,LiKWTM13,CaoKWL12,CaoKWL14,LiDZZ17,LiKD15} is the first of its kind that combines meta-heuristics with a queueing model for VNFPP. We used a simple solution representation where each solution is represented as a string of VNFs and proposed customized mutation and initialization operators to improve the chances of placing at least one instance of each VNF. It also used a simple queuing model that calculates the latency and energy consumption, but it does not consider the packet loss. In this paper, we propose a more advanced solution representation that allows for more diverse solutions without requiring custom genetic operators. Further, we show that this new representation is simple to extend to complex constraints. Last but not the least, we improve upon the model to consider packet loss and show how this significantly affects the quality of solutions.

%% file: problem_formulation.tex


\section{Problem Formulation}
\label{sec:problem_formulation}

This section starts with a descriptive statement of the VNFPP. Then, we give the formal definition and an analysis of the multi-objective VNFPP considered in this paper.

\subsection{Problem Statement}
\label{sec:statement}

\begin{figure}[t!]
	\centering
	\includegraphics[bb=0 0 250 100]{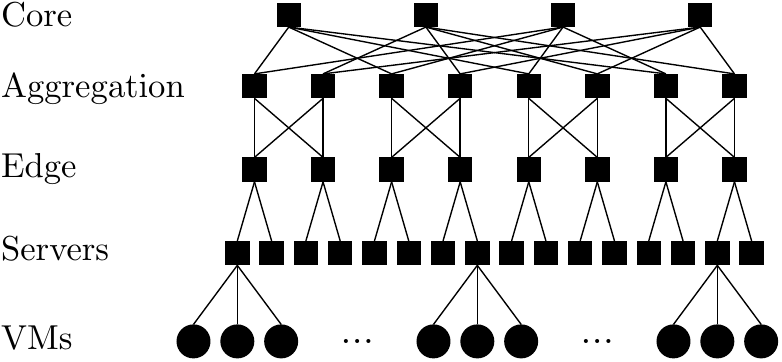}
	\caption{An NFV enabled Fat Tree network topology with $4$ ports and $3$ VMs per server.}
	\label{fig:topology}
\end{figure}

A \underline{data center} consists of a large number servers, each of which can accommodate a limited number of VMs. Traffic is transmitted between servers across the network topology, i.e., a set of switches that interconnect all servers as an example shown in~\pref{fig:topology}. Traffic between VMs on the same server communicate via a virtual switch on the server. In this paper, servers and switches are considered as the data center components that constitute a data center. A \underline{solution to a VNFPP} specifies one or more paths through the network topology for each service. A path is a sequence of data center components that visit each VNF of a service in a prescribed order. Furthermore, each solution also specifies the amount of traffic that should be sent along each path.

The \underline{goal of the VNFPP} is to provide a number of services by placing VNFs on VMs in the data center and defining the paths so as to maximize QoS and minimize capital and operational costs. In this paper, we formulate a three-objective VNFPP that takes two QoS metrics (i.e., latency and packet loss) and a cost metric (i.e., energy consumption) into account.
\begin{itemize}
	\item The \underline{total energy consumption} (denoted as $E_C$).
	\item The \underline{mean latency} of the services (denoted as $L$):
	      \begin{equation}
		      L=\sum_{s\in\mathcal{S}} W_{s}/|\mathcal{S}|,
	      \end{equation}
	      where $W_s$ is the expected latency of $s\in\mathcal{S}$.
	\item The \underline{mean packet loss} of the services (denoted as $P$):
	      \begin{equation}
		      P=\sum_{s\in\mathcal{S}} \mathbb{P}^d_s/|\mathcal{S}|,
	      \end{equation}
	      where $\mathbb{P}^d_s$ is the packet loss probability of $s\in\mathcal{S}$.
\end{itemize}
In particular, $\mathcal{S}$ is the set of services that must be placed and $\mathcal{V}$ is the set of VNFs. A service $s=\{s_1,\cdots,s_n\}\in\mathcal{S}$ is a sequence of VNFs. The network topology is represented as a graph $\mathcal{G}=(\mathcal{C},\mathcal{L})$, where $\mathcal{C}$ denotes the set of data center components and $\mathcal{L}$ denotes the set of links connecting them. A route is a sequence of data center components where $\mathcal{R}^s$ is the set of paths for $s$, $R_{i}^s$ is the $i$-th path of $s$ and $R_{i,j}^s$ is the $j$-th component of this path. The complete notations are listed in Table I of Appendix A\footnote{The appendices can be found in the \href{https://www.dropbox.com/s/urodeyvfdru6807/supp.pdf?dl=0}{supplemental document}.}. There are five constraints associated with this VNFPP. Three of them are core constraints applicable to any VNFPP, and they are defined as follows.
\begin{itemize}
	\item Sequential data center components in a route must be connected by an edge:
	      \begin{equation}
		      (R_{i,j}^s, R_{i,j+1}^s) \in \mathcal{L}.
	      \end{equation}
	\item Each server can accommodate up to $N^V$ VNFs:
	      \begin{equation}
		      \sum_{v\in\mathcal{V}} A_v^{c_{\mathsf{s}}}<N^V,
	      \end{equation}
	      where $A_v^{c_{\mathsf{s}}}$ is the number of instances of the VNF $v$ assigned to the server ${c_{\mathsf{s}}}$.
	\item All VNFs must appear in the route and in the order defined by the service:
	      \begin{equation}
		      \pi^{R^s}_{s_i}\neq\emptyset,\quad\pi^{R_i}_{s_i}<\pi^{R_i}_{s_{i+1}},
	      \end{equation}
	      where $\pi^{R_i}_{s_i}$ is the index of the VNF $s_i$ in the route $R_i$.
\end{itemize}

In practice, security and legal concerns can impose additional constraints.
\begin{itemize}
	\setcounter{enumi}{3}
	\item A business may require an exclusive access to the servers in use due to security or performance restrictions. These requirements can be expressed through \textit{anti-affinity constraints} that restrict which services can share servers. For each service $s\in\mathcal{A}$ where $\mathcal{A}$ is the set of anti-affinity services, the anti-affinity constraints are defined as:
	      \begin{equation}
		      A^{c_{\mathsf{s}}}_{v_1}\cdot A^{c_{\mathsf{s}}}_{v_2}=0,\quad\forall v_1\in s, v_2\notin s.
	      \end{equation}

	\item VNFs may be provided under a license that restricts the number of instances of a VNF that can be created. These are known as the \textit{max instance constraints}:
	      \begin{equation}
		      \sum_{c_s\in\mathcal{C}^s} A_v^{c_{\mathsf{s}}}\leq N_v^M,
		      \label{eq:max_instances}
	      \end{equation}
	      where $N^M_v$ is the maximum number of instances of the VNF $v$ and $\mathcal{C}^{s}$ is the set of all servers.
\end{itemize}

\begin{remark}
    Although these constraints can be considered independently from each other, they are rather complex such that the feasible search space is small relative to the overall search space. Further, the NP-hardness of the problem means that the feasible search space still contains many solutions, yet few of which will be at or near the optimum.
\end{remark}

\begin{remark}
    A multi-objective formulation of the VNFPP helps the decision maker a better understanding of the trade-off between QoS and the energy consumption when increasing the amount of resources spent on services. This also informs how parameters, e.g., network topology and the properties of servers and services, could affect these trade-offs. Last but not the least, it makes an informed selection from the set of possible trade-off solutions.
\end{remark}

\vspace{-1.0em}
\subsection{Analysis of Feasible Search Space}
\label{sec:complexity}

In the context of VNFPP, a solution is feasible if at least one instance of every VNF has been placed. Here we plan to verify that the relative size of the feasible region, which is the probability of a randomly selected solution being feasible, is small. However, due to the NP-hardness of the VNFPP, there is no closed form solution of this relative size. Therefore, we estimate an upper bound instead. In particular, we consider the case where each VNF can be placed at any location independent of whether other VNFs have been placed therein. In the following paragraphs, we first verify that this is indeed an upper bound and then we provide a quantitative estimation to show that it is proportionately small.

\begin{lemma}
	The feasible region under the independence assumption is larger than the exact feasible region.
\end{lemma}

\begin{figure}[t!]
	\centering
	\includegraphics[width=.5\linewidth]{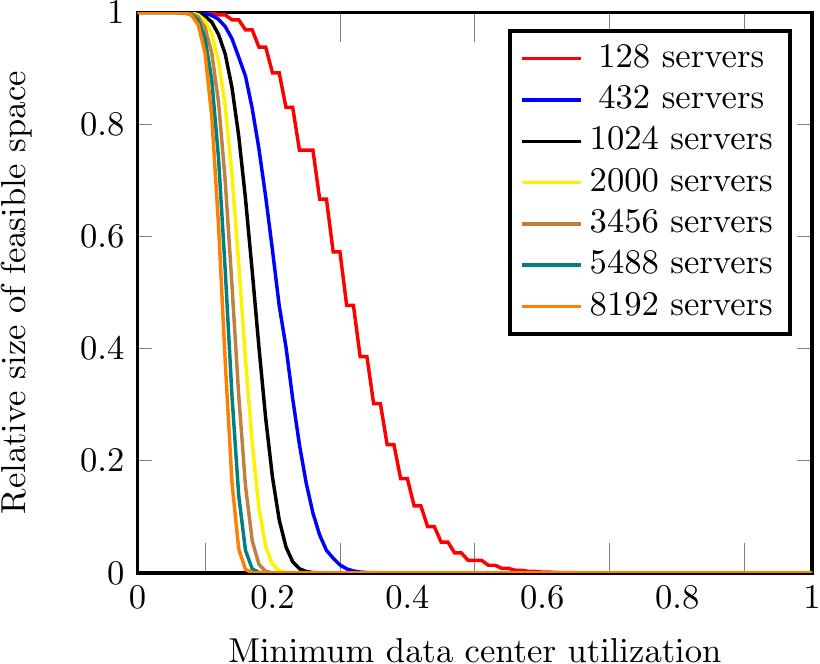}
	\caption{Demonstration of an upper bound of the proportion of the feasible region for different data center sizes.} 
	\label{fig:p_feasible}
	\vspace{1em}
\end{figure}

All proofs can be found in Appendix B. Based on the independence assumption, the probability of a VNF being placed is calculated as:
\begin{equation}
	\mathbb{P}^p_{v}=1-\left(1-\frac{1}{|\mathcal{V}|}\right)^N,
\end{equation}
where $N$ is the number of VMs. Hence, the probability at least one VNF is not placed is calculated as:
\begin{equation}
	\mathbb{P}^{\neg p}=1-\left(\mathbb{P}^p_v\right)^{\mathcal{V}}.
\end{equation}
\pref{fig:p_feasible} plots the probability of generating a feasible solution for a data center with different capacities. These trajectories show that the ratio of the feasible region against the entire search space approaches zero even for very low utilizations. The anti-affinity and max instance constraints further narrow down the feasible region, leading to an increased difficulty.

%% file: system_model.tex
\section{System Model}
\label{sec:system_model}

In this section, we develop an analytical model to derive the three metrics that constitute the objective functions of our VNFPP defined in~\pref{sec:problem_formulation}. They can be calculated for each service by examining the queues in the network. Each data center component consists of one or more buffers where packets are queued before being served. The arrival and service rates at a data center component determine the expected length of each queue, which in turn determine the waiting time and the probability of packet loss as well as the energy consumption. This information can then be used to calculate the latency, packet loss and energy consumption for each path of each service. In the following paragraphs, we first derive the approximation of the arrival rate for each data center component, based on which we calculate the three metrics. 

\subsection{Arrival Rates of Data Center Components}
\label{sec:arrival_rate}

To calculate the arrival rate we must establish some reasonable assumptions about the system's behavior. In line with \cite{PoissonTraffic}, we assume the traffic generated by end users follows a Poisson distribution with a mean rate $\lambda_s$. As end users access the service independently, the total traffic arrival rate of a service can be calculated as the superposition of multiple independent Poisson processes. When packets arrive at a data center component, they are served with a first-in-first-out queueing strategy. To make the analytical model applicable to the practical implementation, instead of exploiting the infinite queueing strategies in \cite{InfiniteQueue}, we assume each data center component has a finite buffer length $B_c$. If the buffer becomes full, the newly arrived packets would be dropped to avoid system congestion. Finally, since packets are processed independently, the time for a data center component to process a packet follows an exponential distribution with service rate $\mu_c$. Under these conditions, we model the service processing at each data center component as an $M/M/1/B_c$ system.

\begin{figure}[t!]
	\centering
	\begin{minipage}{0.5\textwidth}
		\centering
		\resizebox{0.5\textwidth}{!}{
			\includegraphics{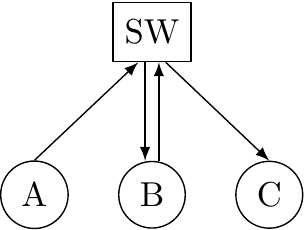}
		}
		\subcaption{Original Configuration}
	\end{minipage}\hfill
	\begin{minipage}{0.5\textwidth}
		\centering
		\resizebox{0.5\textwidth}{!}{
			\includegraphics{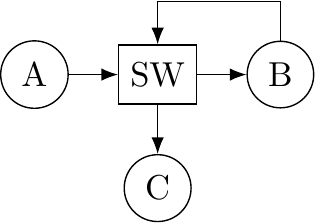}
		}
		\subcaption{Unfolded Loop}
	\end{minipage}
	\vspace{.2em}
	\caption{Three VNFs (A, B, C) are visited in sequence through a single switch (SW). This forms a loop causing the arrival rate at the switch to be dependent on its own packet loss.}
	\label{fig:arrival_loop}
\end{figure}

Next we can calculate the arrival rate of each data center component. Let $\lambda_c$ be the arrival rate of a data center component $c\in\mathcal{C}$. It is the sum of the packet flow rates of all paths entering this data center component. Due to the finite buffer size, the effective arrival rate $\lambda_c^e$ is less than the arrival rate and calculated as $\lambda _{ c }^{ e }=\lambda _{ c }{ \left( 1-\mathbb{P}^d_{c} \right)  }$, where $\mathbb{P}^d_{c}$ is the packet loss probability and is calculated as~\cite{Kleinrock75}:
\begin{equation}
    \mathbb{P}^d_{c}
	=
	\begin{cases}
		\frac{(1-\rho)\rho^{B_c}}{1-\rho^{B_c+1}}, & \text{if}\ \lambda\neq\mu\\
        \frac{1}{B_{c}+1}, & \text{otherwise}
	\end{cases},
	\label{eq:pl}
\end{equation}
where $\rho=\lambda_{c}/\mu_{c}$. If the packet loss at a data center component were fixed then the arrival rate at each component would simply be the sum of the packet flow rates of the routes through that component. In practice, since the packet loss at a data center component depends on the arrival rate at the earlier components on the same path, dependency loops can form if the same component is visited multiple times in a sequence (as demonstrated in~\pref{fig:arrival_loop}). In this case, the packet loss probability at the revisited component becomes a function of its own arrival rate thus resulting in a dynamic system. Since the arrival rate at each component changes over time in a dynamic system, it is significantly more complex to derive the performance metrics based on the arrival rate. Existing works unfortunately neglect this factor by either considering models without packet loss (e.g., \cite{ZhangXLLGW17,QuZYSLR20,AgarwalMCD18}) or simply ignoring the dynamic feature but only calculating the arrival rate at the outset instead (e.g.,~\cite{ChuaWZSH16,MarottaZDK17}). This paper proposes an iterative method to calculate the expected arrival rate over time. We first show that the arrival rates at all data center components naturally converge towards a fixed point given infinite time. Then, we elaborate the method that derives the expected arrival rate.

\begin{lemma}
The arrival rate at each data center component converges towards a fixed point as time approach infinity.
\label{lemma:arrival_rate}
\end{lemma}

Based on~\pref{lemma:arrival_rate}, a na\"ive method to calculate the arrival rate is to evaluate the upper and lower bounds of the arrival rate until they converge. However, this method is impractical since this requires infinite time. Instead, we propose to approximate the arrival rate by calculating the bounds until they converge to the point that further iterations are unlikely to change the expected arrival rate more than a threshold $\delta>0$. As the pseudo-code shown in Algorithm 2 in the Appendix C, it first initializes the packet loss at each data center component to zero to simulate there being no packet in any queue (lines 3 to 4). In the main loop, the algorithm first calculates the current arrival rate and packet loss for each data center component by using the previous settings of packet loss (lines 6 to 8). From~\pref{lemma:arrival_rate}, we can see that the current arrival rate will be either a lower or upper bound of the arrival rate. Next, the algorithm calculates the mean of the upper and lower bounds of the arrival rate for each data center component (line 12) and the divergence from the previous mean for each component (line 14). If the maximum divergence from the mean across all components has remained below $\delta$ for $\gamma>1$ iterations, future iterations are unlikely to alter the mean arrival rate. We thus terminate the process and output the mean arrival rate as the arrival rate at each data center component (lines 17 to 24).

Note that the parameters $\delta$ and $\gamma$ determine the accuracy and convergence speed of this model. A lower $\delta$ increases the model accuracy by requiring the mean to be more stable before being considered converged. The model is less sensitive to $\gamma$ which is required for the rare scenario where the bounds temporarily appear converged. We found that $\delta = 5.0$ and $\gamma = 10$ give a balanced trade-off between efficiency and accuracy. 

\subsection{Service Packet Loss}
\label{sec:packet_loss}

The packet loss probability of a service is the expected packet loss considering the probability of selecting each path:
\begin{equation}
    \mathbb{P}^d_s=\sum_{i=1}^{|\mathcal{R}^s|} \mathbb{P}^d_{R^s_i}\cdot \mathbb{P}_{R^s_i},
	\label{eq:pl_service}
\end{equation}
where $\mathbb{P}^d_{R^s_i}$ is the probability that a packet is dropped at any component on the path $R^s_i$. It is calculated as:
\begin{equation}
	\mathbb{P}^d_{R^s_i}=1-\prod_{c\in R^s_i}\left(1-\mathbb{P}^d_c\right).
	\label{eq:pl_path}
\end{equation}

\subsection{End-to-End Latency}
The end-to-end latency of a service is the expected waiting time over all paths. It is calculated as:
\begin{equation}
	W_s=\sum_{i=1}^{|\mathcal{R}^s|} W_{R^s_i} \cdot\mathbb{P}_{R^s_i},
\end{equation}

\noindent where $W_{R^s_i}$ is the average latency for $R^s_i$ and it is calculated as the sum of the waiting time at each data center component:
\begin{equation}
	W_{R^s_i} = \sum_{c\in R^s_i} W_c,
\end{equation}
where $W_c=\overline{N}/\hat{\lambda}_c$ is the waiting time at the component $c\in R^s_i$ and $\hat{\lambda}_c=\lambda_c\cdot\left(1-\mathbb{P}^d_c \right)$ is its effective arrival rate and $\overline{N}$ is its expected queue length~\cite{Kleinrock75}:
\begin{equation}
	\overline{N} = \begin{cases}
		\frac{\rho[1 - (B_c + 1)\rho^{B_c} + B_c\rho^{B_c+1}]}{(1 - \rho)(1 - \rho^{B_c+1})}, & \text{if } \ \lambda \neq \mu \\
		B_c/2,                                                                      & \text{otherwise}
	\end{cases}.
\end{equation}

\subsection{Energy Consumption}
\label{sec:energy}

The total energy consumption of a data center is the sum of energy consumed by each of its components. The energy consumption process follows a three-state model with \texttt{off}, \texttt{idle} and \texttt{active} states. Specifically, a component is \texttt{off} if its arrival rate is zero; it is \texttt{idle} while it is not processing any packet; otherwise, the component is \texttt{active}. A data center component does not consume any energy when it is \texttt{off}. Thus, we only need to consider the energy consumption of its \texttt{active} and \texttt{idle} states, denoted as $E^A$ and $E^I$, respectively. The total energy consumption of a data center is the sum of energy consumed by all its components:
\begin{equation}
    E_C=\sum_{c\in\mathcal{C}\setminus\mathcal{C}^{\mathsf{vm}}} U_c\cdot E^A+(1-U_c)\cdot E^I,
	\label{eq:sum_energy}
\end{equation}
where $\mathcal{C}^{\mathsf{vm}}$ is the set of VMs and $U_c$ is the utilization of the data center component $c$. To calculate $U_c$, we need to consider both single- and multiple-queue devices. The utilization of a queue is given by:
\begin{equation}
	\overline{U}_c =
	\begin{cases}
		0, & \rm{if} \ \lambda=0 \\
		\frac{1-\rho}{1-\rho^{B_c+1}}, & \rm{if}\ \lambda\le\mu \\ 
		\frac{1}{B_c+1}, & { \rm{otherwise} }
	\end{cases}.
	\label{eq:u}
\end{equation}
Physical switches can be modeled with a single-queue for their buffers. Hence the utilization of a switch $U_c$ is equal to the utilization of its queue:
\begin{equation}
    U_{c \in\mathcal{C}^{\mathsf{sw}}} = \overline{U}_c,
\end{equation}
where $\mathcal{C}^{\mathsf{sw}}$ is the set of switches. A server has multiple buffers: one for the virtual switch and one for each VNF. The server is \texttt{idle} when no packets are being processed at any of its buffers. Thus, the utilization of a server is calculated as:
\begin{equation}
	U_{c_{\mathsf{sr}}\in\mathcal{C}^\mathsf{sr}}=1-\left(1-\overline{U}_{c_{\mathsf{vs}}} \right)\cdot\prod_{c_{\mathsf{v}}\in\mathcal{A}^{c_{sr}}}\left(1-\overline{U}_{c_{\mathsf{v}}} \right),
	\label{eq:us}
\end{equation}
where $\mathcal{C}^\mathsf{sr}$ is the set of servers, $c_{\mathsf{vs}}$ is the virtual switch of the server and $\mathcal{A}^{c_\mathsf{sr}}$ is the set of VNFs assigned to the server $c_\mathsf{sr}$.

%% file: optimisation.tex

\section{Proposed Evolutionary Optimization Framework for VNFPPs}
\label{sec:optimisation}

In this section, we propose a tailored evolutionary optimization framework to solve the VNFPP defined in~\pref{sec:problem_formulation}. There are two tailored features: one is a genotype-phenotype solution representation, detailed in~\pref{sec:representation}, that guarantees feasible solutions for the underlying VNFPP; the other is a tailored initialization operator, detailed in~\pref{sec:operators}, built upon the solution representation to produce a promising initial population. Note that these tailored features can be readily incorporated into any existing EMO algorithm as shown in~\pref{sec:incorporation}. Although these operators are specifically designed for the popular Fat Tree network topology~\cite{AlFaresLV08} (see~\pref{fig:topology}), we argue that our model and problem formulation are generally useful for any network topology.

\subsection{Genotype-Phenotype Solution Representation}
\label{sec:representation}

\begin{figure*}[t!]
	\centering
	\includegraphics[width=\linewidth]{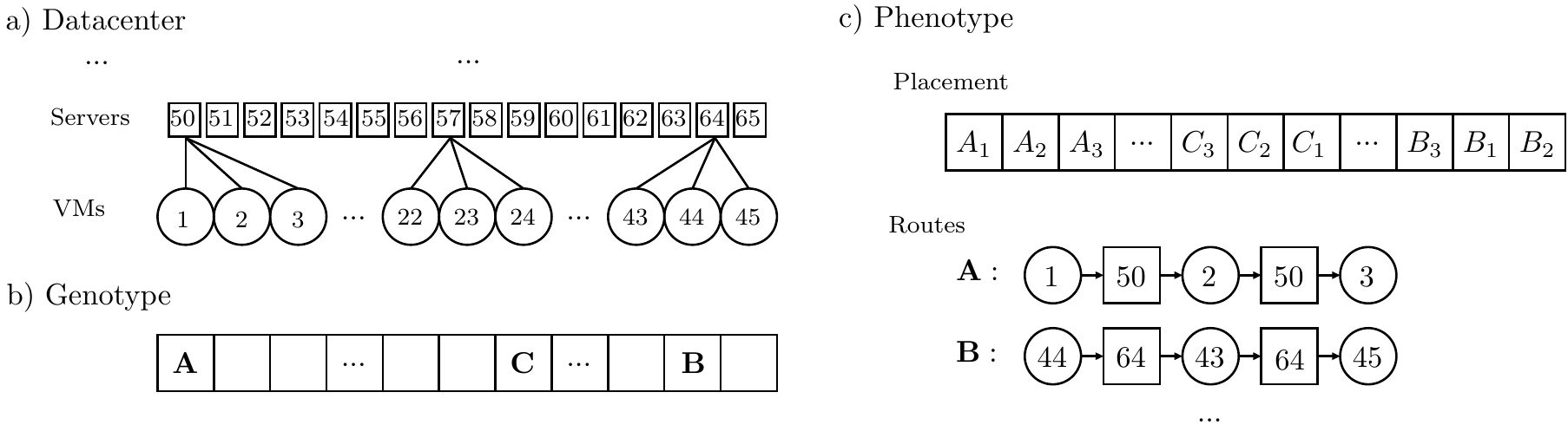}
	\caption{An illustrative example of the genotype-phenotype solution representation developed in this work.}
	\label{fig:gp_mapping}
\end{figure*}

One of the key challenges when designing and/or applying EAs to real-world optimization problems is how to encode the problem into a solution in EA. In this paper, we propose a genotype-phenotype solution representation for our VNFPP. As shown in \pref{fig:gp_mapping}, the genotype is a string of characters where each character can be either a service $s\in S$ or a sentinel \texttt{NONE}, i.e., the corresponding VM is not in use (illustrated in the figure with an empty character). The phenotype is a set of paths and the corresponding path probabilities required for the VNFPP. The mapping between them defines how to transform the genotype into the corresponding phenotype. Due to the existence of complex constraints defined in~\pref{sec:problem_formulation}, a simple mapping does not always lead to a feasible solution. The main crux of our genotype-phenotype solution representation is the use of problem-specific heuristics at the mapping stage that avoid generating infeasible solutions. It consists of three steps: \texttt{balance}, \texttt{placement} and \texttt{routing}.

\begin{figure*}[t!]
	\begin{subfigure}[T]{.3\linewidth}
		\centering
		\includegraphics[width=\columnwidth]{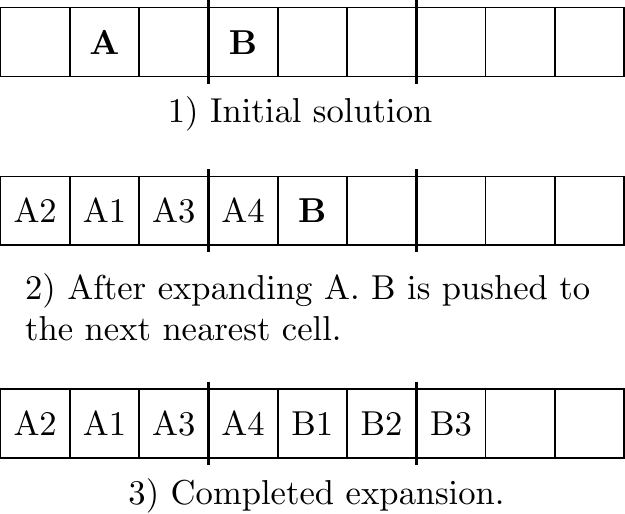}
		\caption{Simple expansion.}
		\label{fig:1a}
	\end{subfigure}\hfil
	\begin{subfigure}[T]{.3\linewidth}
		\centering
		\includegraphics[width=\columnwidth]{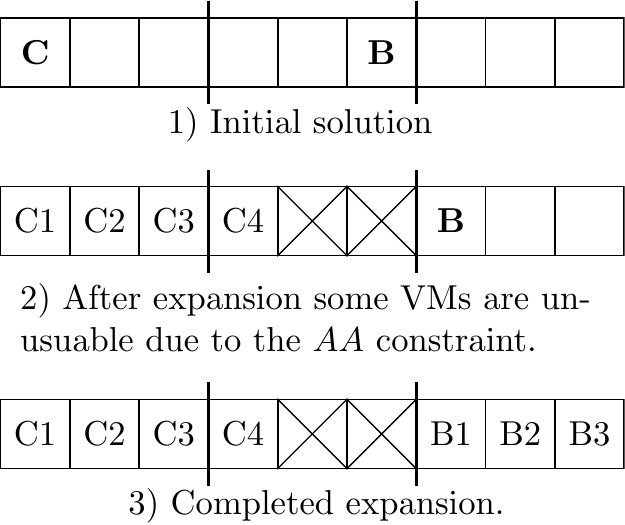}
		\caption{Anti-affinity expansion.}
		\label{fig:1b}
	\end{subfigure}\hfil
	\begin{subfigure}[T]{.3\linewidth}
		\centering
		\includegraphics[width=\linewidth]{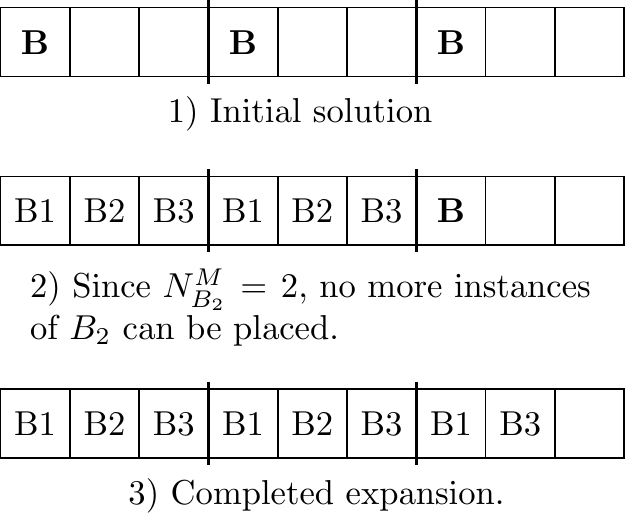}
		\caption{Max instances expansion}
		\label{fig:1c}
	\end{subfigure}
	\vspace{2em}
	\caption{This figures shows three examples of the placement step of the genotype-phenotype mapping. There are three services here: service $\mathbf{A} = \{ A_1, A_2, A_3, A_4\}$ , service $\mathbf{B} = \{B_1, B_2, B_3\}$ which has a max instances constraint $N_{B_2}^M = 2$ and service $\mathbf{C} = \{C_1, C_2, C_3, C_4\}$. In addition, $\mathcal{A} = \{C\}$ such that service $\mathbf{C}$ cannot share a server with any other service. Tall lines indicate server boundaries.}
	\label{fig:placement}
\end{figure*}

\subsubsection{Balance}
The \texttt{balance} step adds and/or removes service instances to guarantee the feasibility after the genotype-phenotype mapping. This is implemented by ensuring that the genotype has at least one instance of each service and the total number of VMs being used does not exceed the available number in the data center. The pseudo-code is given in Algorithm 3 in Appendix C. It first identifies the location of all unused VMs along with the location and number of each service instance (lines 6 to 14). Using this information, the algorithm can calculate the number of VMs the solution will require after the mapping (lines 16 to 23) and identify missing services that have no service instances (lines 24 to 28). The algorithm then places a service instance for any missing services on a free VM if possible (lines 29 to 33). Finally, if there is insufficient space to place a missing service instance or the expanded length of the solution would exceed the total capacity, the algorithm removes the service instance with the lowest contribution and, if necessary, replaces it with a service instance for a missing one (lines 35 to 43). In particular, the contribution of an instance is evaluated as the change in the service instance utilization if it were removed:
\begin{equation}
	C^s_i = \frac{\lambda_{s}}{\mu_{s_1} \cdot (i - 1)} - \frac{\lambda_{s}}{\mu_{s_1} \cdot i},
	\label{eq:contribution}
\end{equation}
\noindent where $C^s_i$ is the contribution of the $i$th instance of $s$ and $\mu_{s_1}$ is its service rate of the first VNF. As the arrival rate is distributed over each VNF, a service with several instances will have some instances with a low contribution. On the other hand, if a service has only one instance, it will have an infinite contribution. This minimizes the impact on the service quality when removing solutions. 

\subsubsection{Placement}
The \texttt{placement} step uses a first feasible heuristic, a variant of the first fit heuristic from the cloud computing literature~\cite{KellerTLB12}, to place the VNFs of a service in the phenotype close to the position of the service instance in the genotype without violating any constraint. The first feasible heuristic is executed on each service instance. It places the first VNF of the service on the nearest VM to the service instance that would not result in a constraint violation. This is repeated from the new position for the next VNF instance until all VNFs are placed. As anti-affinity services reserve the whole of a server, they must be placed first to ensure the service is not fragmented across multiple servers and does not reserve more space than necessary. \pref{fig:placement} presents three examples of the \texttt{placement} step for different scenarios.

\subsubsection{Routing}
Finally, the \texttt{routing} step finds the set of shortest paths between the VNFs of each service instance to complete a solution. A Fat Tree network can be efficiently traversed by stepping upwards to the parent switch until a common ancestor between the initial and the target VNFs is found. In the Fat Tree network topology, there can be several routes between VNFs sharing the same distance. In this paper, we apply the equal-cost multi-path routing strategy~\cite{Hopps2000} to distribute the traffic evenly over all shortest paths between sequential VNFs. This strategy has been shown to be optimal for Clos data center networks such as the Fat Tree network~\cite{ChiesaKS17}.

\vspace{-.8em}
\subsection{Tailored Initialization Operator}
\label{sec:operators}


The goal of initialization is to generate a set of diverse initial solutions to \lq jump start\rq\ the search process afterwards. Note that both the placement and the number of instances in the VNFPP can influence the solution quality. Uniform sampling, one of the most popular initialization strategies, varies the placement of service instances but the expected number of instances remains the same across all solutions. To amend this problem, we propose a variant of uniform sampling where service instances are placed uniformly at random, but the number of instances of each service varies across the population. More specifically, we first calculate the maximum number of instances of each service that can be accommodated in a data center. Thereafter, the solution is initialized by placing some fraction of this number of instances of each service. For the $i$th solution, the number of instances of the service $s$ is calculated as:
\begin{equation}
	N^I_{i,s} = \left\lfloor \frac{i}{N} \cdot \frac{N}{\sum_{s\in S} \abs{s}} \right\rfloor,
	\label{eq:num_services}
\end{equation}
where $N$ is the population size. For example, if the population size $N=100$, the $100$th solution will have twice as many instances of each service as the $50$th solution.

\subsection{Incorporation into EMO Algorithms}
\label{sec:incorporation}

The solution representation and initialization operators proposed in~\pref{sec:representation} and~\pref{sec:operators} can be incorporated into any EMO algorithm~\cite{LiKWCR12,LiWKC13,CaoKWLLK15,LiDY18,WuKZLWL15,LiKCLZS12,LiDAY17,LiDZ15,LiXT19,GaoNL19,LiuLC19,LiZ19,KumarBCLB18,CaoWKL11,LiX0WT20,LiuLC20,LiXCT20,WangYLK21,ShanL21,LaiL021,LiLLM21,WuKJLZ17,LiCSY19,LiLDMY20,WuLKZ20,PruvostDLL020,XuLA22,LiLL22,ZhouLM22,ChenLTL22,Williams0M22,FanLT20}. In this paper, we consider three iconic EMO algorithms including NSGA-II~\cite{DebAPM02}, IBEA~\cite{ZitzlerK04}, and MOEA/D~\cite{ZhangL07} for a proof-of-concept purpose. Note that we do not need to make any modification on the environmental selection of the baseline algorithm. Further, we find the classic uniform crossover and mutation reproduction operators are already sufficient to vary and exchange information on the number and position of service instances.

%% file: experiments.tex

\section{Empirical Study}
\label{sec:experiments}

We seek to answer the following six research questions (RQs) through our empirical study.
\begin{enumerate}[RQ1:]
    \item How accurate is the QoS model developed in~\pref{sec:system_model} compared to the SOTA in the literature?
    \item How is the performance different EMO frameworks on the VNFPP?
    \item Is an accurate QoS model beneficial compared to the widely used surrogate models in the literature?
    \item Does the proposed solution representation improve upon alternative solution representations?
    \item How does the tailored EMO algorithm compare against the SOTA peer algorithms?
    \item How well do the proposed operators cope with challenging constraints in VNFPP?
\end{enumerate}

\subsection{Experimental Settings}
\label{sec:settings}

The parameter settings used in this work are listed in Table II in the Appendix D. To reflect the mechanism of real switches, the service rate and queue length of each switch in our model are the sum of the service rates and queue lengths of each port, e.g., a switch with $8$ ports will have a service rate of $8\times 20=160$ requests per ms. To create a VNFPP instance, we generate enough services to hit the target minimum data center utilization, e.g., if the data center has $1,000$ VMs, the expected service length is $5$ and the target data center utilization is $50$\%, then there will be $100$ distinct services. Next, the service arrival rate and length and the VNF service rate are set for each service and VNF by sampling from a Gaussian distribution whose specification is given Table II in the Appendix D. The quality of these non-dominated solutions is evaluated by the Hypervolume (HV) metric~\cite{ZitzlerT99} that measures both the convergence and diversity of the population, simultaneously.

\subsection{Evaluation of QoS Model Accuracy}
\label{sec:model_accuracy}

\subsubsection{Methods}
The QoS model developed in~\pref{sec:system_model} stands for the foundation of this study. Its correctness and accuracy determine whether our algorithm is applicable to real-world scenarios. To answer RQ1, we evaluate the accuracy of the model by comparing its predictions against benchmark measurements taken from a simulated data center.

Specifically, we generate $100$ VNFPP instances for a data center with $412$ servers to constitute a diverse benchmark suite. Then, we use our proposed initialization operator (see \pref{sec:operators}) to generate $100$ candidate solutions for each VNFPP instance. Next, we evaluate each solution by using our proposed model and select four solutions for evaluation: 1) the one with the lowest latency; 2) the one with the lowest packet loss; 3) the one with the lowest energy consumption; and 4) the one that best balances all objectives. By using diverse solutions, we can rule out any inaccuracy reflected by the model. For example, if the model is poor at predicting the latency, the data set will contain a solution with a high expected latency that highlights this issue. 

To get accurate measurements for the benchmark, we apply a discrete event simulator (DES) to calculate each metric of a solution. A DES simulates the transmission of each packet through the data center to produce accurate measurements of the QoS and energy consumption. In our experiments, the DES is based on the same assumptions introduced in \pref{sec:system_model} and it is used to evaluate each solution for a range of arrival rates.

We compare our model against two other accurate models used in the literature.
\begin{itemize}
    \item \underline{$M/M/1$ queueing model}: As one of the most popular models in the literature~\cite{PeiHXLWW20,JemaaPP16,BaumgartnerRB15}, it models the data center as a network of queues and assumes that each queue has an infinite length. Under this assumption, there is no packet loss. However, if the arrival rate at a queue is greater than or equal to its service rate, the length of the queue will go to infinity that leads the waiting time to approach infinity and the utilization to approach $100$\%.
    \item \underline{$M/M/1/B_c$ queuing model}: In contrast, this model consider queues with a finite length. Existing $M/M/1/B_c$ queueing models like~\cite{ChuaWZSH16} consider packet loss but not feedback loops. In essence, they calculate the instantaneous arrival rate and packet loss at each data center component when the services are first started.
\end{itemize}

\subsubsection{Results}
\begin{figure*}[t!]
    \centering
    \includegraphics[width=\textwidth]{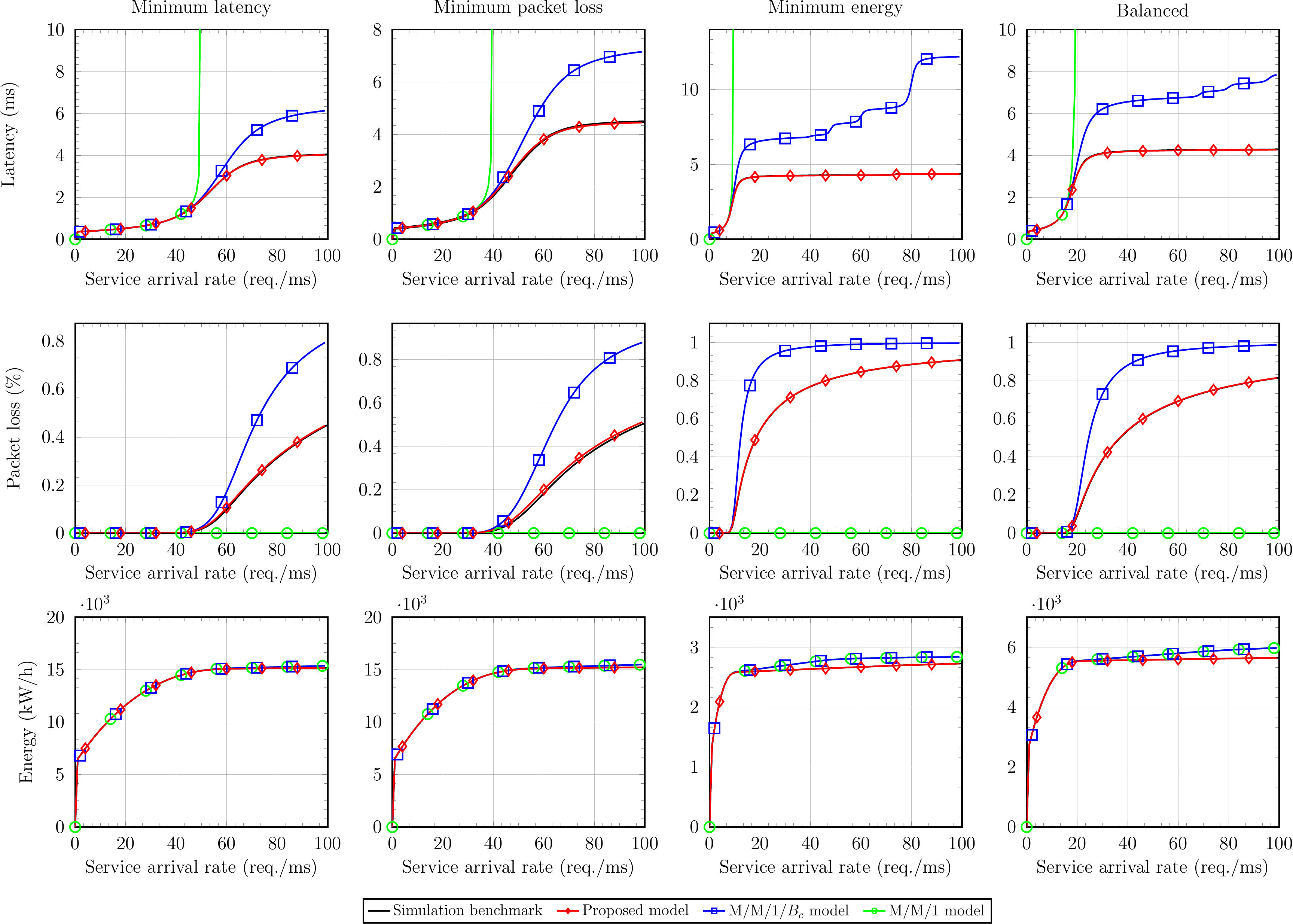}
    \caption{Comparison results of three different metrics estimated by our proposed model against the other two queueing models.}
    \label{fig:model_sim}
\end{figure*}

\pref{fig:model_sim} shows the estimates of each metric by different models benchmarked against the ground-truth measurements. The closer the model matches the benchmark, the more accurate the model is. From the results, we find that our proposed model is significantly more accurate than the other peer queueing models. This reflects how our model correctly captures the impact of feedback loops on the QoS and energy consumption.

In contrast, the $M/M/1/B_c$ queueing model, which does not account for feedback loops, is overly pessimistic with a high latency, packet loss and energy consumption. This is because the $M/M/1/B_c$ model calculates the instantaneous arrival rate at the start of operation. As shown in Lemma \ref{lemma:arrival_rate}, this is always higher than the arrival rate after convergence.

Likewise, these results also demonstrate the drawbacks of the commonly used $M/M/1$ model. First, the model falsely assumes that the queue at a component can grow to an infinite length and as a consequence it believes that the packet loss is $0$ in all scenarios. As a result, the model becomes less reliable as the arrival rate increases and the packet loss becomes large.

\begin{tcolorbox}[sharpish corners, top=2pt, bottom=2pt, left=4pt, right=4pt, boxrule=0.0pt, colback=black!5!white,leftrule=0.75mm,]
    \textbf{\underline{Response to RQ1:}} \textit{Due to an understanding of the impact of feedback loops, our proposed model gives significantly more accurate estimates of the QoS and energy consumption than other models when the packet loss is considered.}
\end{tcolorbox}
\vspace{-1.0em}

\subsection{Comparison with Other EMO Frameworks}
\label{sec:moea_comparison}

\subsubsection{Methods}
To validate that our operators can be integrated into any EMO framework, we compare the quality of solutions obtained when different EMO algorithms developed in~\pref{sec:incorporation}. In our experiments, we generate $30$ VNFPP instances for six data centers with different sizes. To compare the performance of different algorithms, we use the QoS model developed in~\pref{sec:system_model} to evaluate the objective functions of the solutions obtained by different algorithms and use the HV indicator as the performance measure.

\subsubsection{Results}
\begin{figure}[t!]
    \centering
    \includegraphics[width=.5\linewidth]{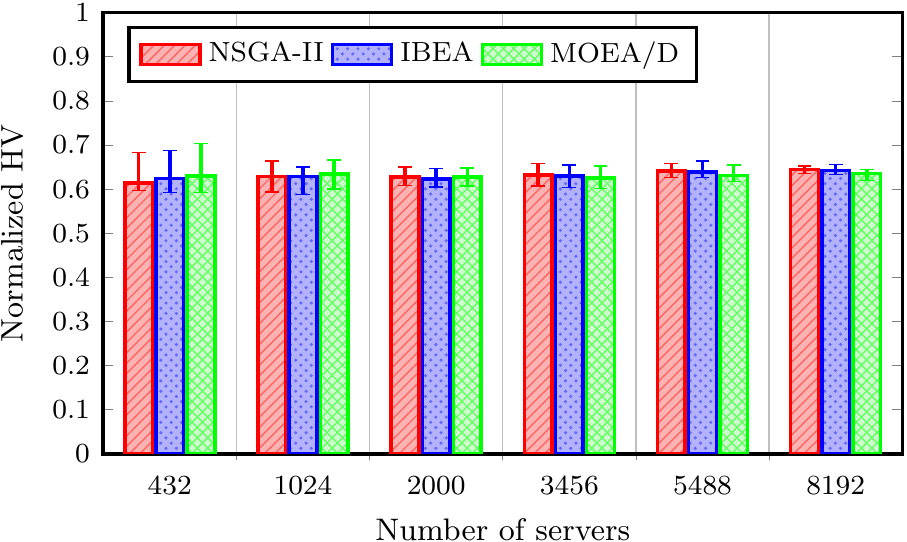}
    \caption{The lower, median, and upper quartiles of HV obtained by different EMO algorithms on $30$ VNFPP instances.}
    \label{fig:moea_comparison}
\end{figure}

The results of this test are illustrated in~\pref{fig:moea_comparison}. We found that all algorithm performed similarly, with no algorithm performing consistently significantly better than any other. Given all algorithms performed similarly, we have selected NSGA-II for use in future tests based on its widespread adoption in the literature. 

\begin{tcolorbox}[sharpish corners, top=2pt, bottom=2pt, left=4pt, right=4pt, boxrule=0.0pt, colback=black!5!white,leftrule=0.75mm,]
    \textbf{\underline{Response to RQ2:}} \textit{From our empirical results, we find that all EMO algorithms using our proposed operators have shown comparable performance.}
\end{tcolorbox}
\vspace{-1.0em}

\subsection{Benefit of Queueing Models}
\label{sec:model_benefit}

\subsubsection{Methods}
To answer RQ3, we compare the quality of solutions obtained by our tailored NSGA-II when using different QoS models including three queueing models studied in~\pref{sec:model_accuracy} and three popular surrogate models briefly introduced as follows.
\begin{itemize}
    \item\underline{Constant waiting time or packet loss (CWTPL)}: As discussed in~\cite{HawiloJS19} and~\cite{VizarretaCMMK17}, this model assumes a constant waiting time at each data center component. In addition, we also keeps the packet loss probability at each component as a constant. Based on these assumptions, we can evaluate the latency and packet loss for each service and apply the metric of the energy consumption developed in~\pref{sec:energy}. All these constitute a three-objective problem that aims to minimize the average latency, packet loss and total energy consumption.

    \item\underline{Resource utilization (RU)}: As in~\cite{ChantreF20,QiSW19} and~\cite{GuoWLQA0Y20}, this model assumes that the waiting time is a function of the CPU demand and the CPU capacity of each VM. In addition, the demand is assumed to determine the packet loss probability as well. Based on these assumptions, we evaluate the latency for each service and apply the metric of the energy consumption developed in~\pref{sec:energy}. All these constitute a two-objective problem that aims to minimize the average latency (and by extension the packet loss) and the total energy consumption.

    \item\underline{Path length and used servers (PLUS)}: This model uses the percentage of used servers to measure the energy consumption (e.g.,~\cite{MiottoLCG19,RankothgeLRL17,LiuZDLGZ18}) and the length of routes for each service as a measure of service latency, packet loss or quality (e.g.,~\cite{LuizelliCBG17,AllegKMA17,BeckB15}). All these constitute a two-objective problem that aims to minimize the path length and the number of used servers.
\end{itemize}

In our experiments, we generate $30$ problem instances of the Fat Tree data center with $432$, $1,024$, $2,000$, $3,456$, $5,488$, and $8,192$ servers respectively. At the end, the non-dominated solutions found by our tailored NSGA-II with different QoS models are re-evaluated by using the QoS model developed in~\pref{sec:system_model}. 

\subsubsection{Results}
\begin{figure*}[t!]
    \centering
    \includegraphics[width=\linewidth]{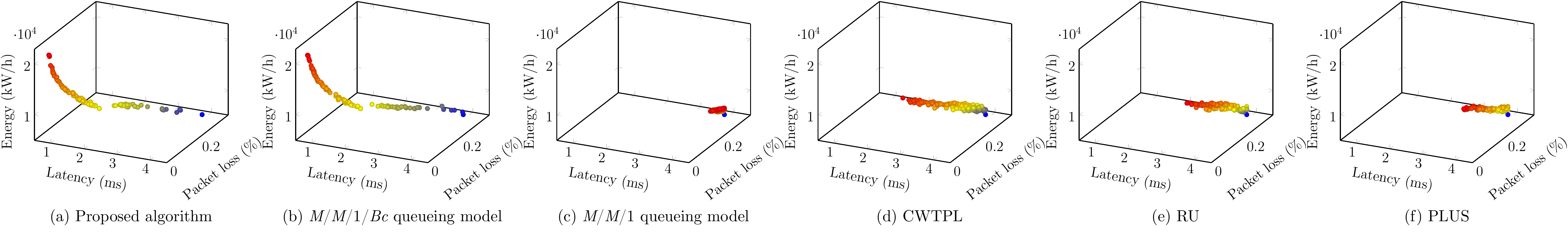}
    \caption{Non-dominated solutions obtained by NSGA-II using our proposed model and models from the literature. More diverse solutions with lower objective values indicate more appropriate models.}
    \label{fig:model_objectives}
\end{figure*}

\begin{figure}[t!]
    \centering
    \includegraphics[width=.6\linewidth]{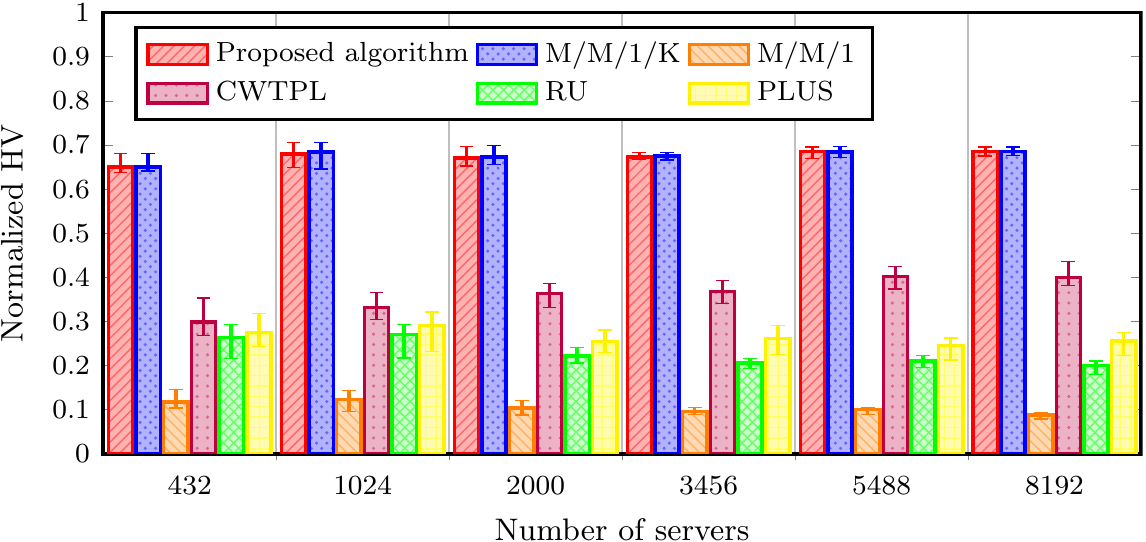}
    \caption{The lower quartile, median, and upper quartile of the HV of the population obtained by using different surrogate models to evaluate the objective functions of our VNFPP.}
    \label{fig:model_benefits}
\end{figure}

From the results shown in Figs.~\ref{fig:model_objectives} and~\ref{fig:model_benefits}, we find that the solutions obtained by using our proposed model and the $M/M/1/B_c$ queueing model are comparable with each other while they are significantly better than those obtained by using other models in terms of the population diversity.


Specifically, populations obtained with the $M/M/1$ queueing model have a poor diversity. This can be attributed to the inability of the $M/M/1$ queueing model to distinguish solutions by the latency or packet loss metrics. In particular, most solutions obtained by using the $M/M/1$ queueing model have an infinite latency and no packet loss. This is because if the arrival rate at any data center component is larger than the service rate, the waiting time at that component tends towards infinity. Hence the average latency also tends towards infinity. As all solutions have the same latency and packet loss, NSGA-II can only distinguish solutions based on their energy consumption. As a result, only solutions with low energy consumption survive.

For a similar reason, the surrogate models also failed to produce diverse solutions. Despite their differences, none of the surrogate models provide any incentive to vary the number of service instances. For example, both the CWTPL and PLUS models benefit from shorter average routes. However, increasing the number of service instances will also increase the number of servers yet is unlikely to decrease the average service length. Likewise, for the RU model, increasing the number of service instances levels up the energy consumption and makes it more difficult for the algorithm to find servers with a low resource utilization.

\begin{tcolorbox}[sharpish corners, top=2pt, bottom=2pt, left=4pt, right=4pt, boxrule=0.0pt, colback=black!5!white,leftrule=0.75mm,]
    \textbf{\underline{Response to RQ3:}} \textit{The queueing models that account for the packet loss lead to significantly more diverse solutions compared to the other queueing model(s) as well as the surrogate models.}
\end{tcolorbox}
\vspace{-1.5em}

\subsection{Evaluation of Solution Representation}
\label{sec:alternative_representations}

To answer RQ4, we compare the quality of solutions obtained when different solution representations are applied to the VNFPP. Specifically, the following two meta-heuristic algorithms use NSGA-II as the baseline but have different solution representations.
\begin{itemize}
    \item\underline{Binary representation}: As in \cite{ChantreF20,KaurGK020} and~\cite{CharismiadisTPM20}, a string of binary numbers are used to represent if a VNF is assigned to a server.
    \item\underline{Direct representation}: As in \cite{RankothgeLRL17}, a solution is directly represented as a string of VNFs.
\end{itemize}
In our experiments, we generate $30$ VNFPP instances for six data centers with different sizes. To compare the performance of different algorithms, we use the QoS model developed in~\pref{sec:system_model} to evaluate the objective functions of the solutions obtained by different algorithms.

\begin{figure}[t!]
    \centering
    \includegraphics[width=.6\linewidth]{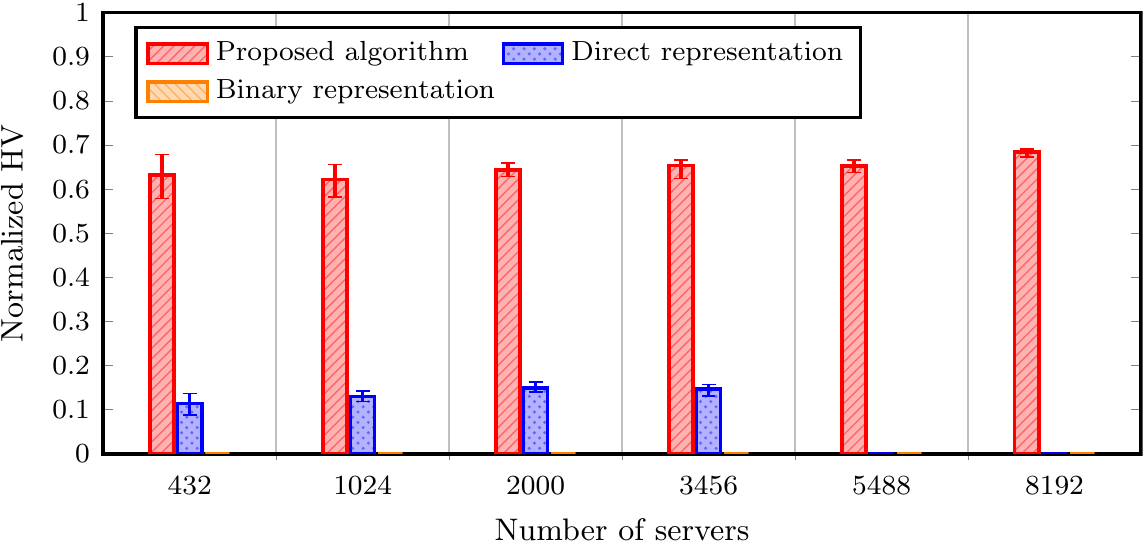}
    \caption{The lower quartile, median, and upper quartile of the hyper-volume of the population for different algorithms on $30$ VNFPP instances found by NSGA-II using different solution representations on $30$ VNFPP instances.}
    \label{fig:solution_representation_comparison}
\end{figure}
\begin{figure}[t!]
    \centering
    \begin{subfigure}[b]{0.4\linewidth}
        \includegraphics[width=.8\textwidth]{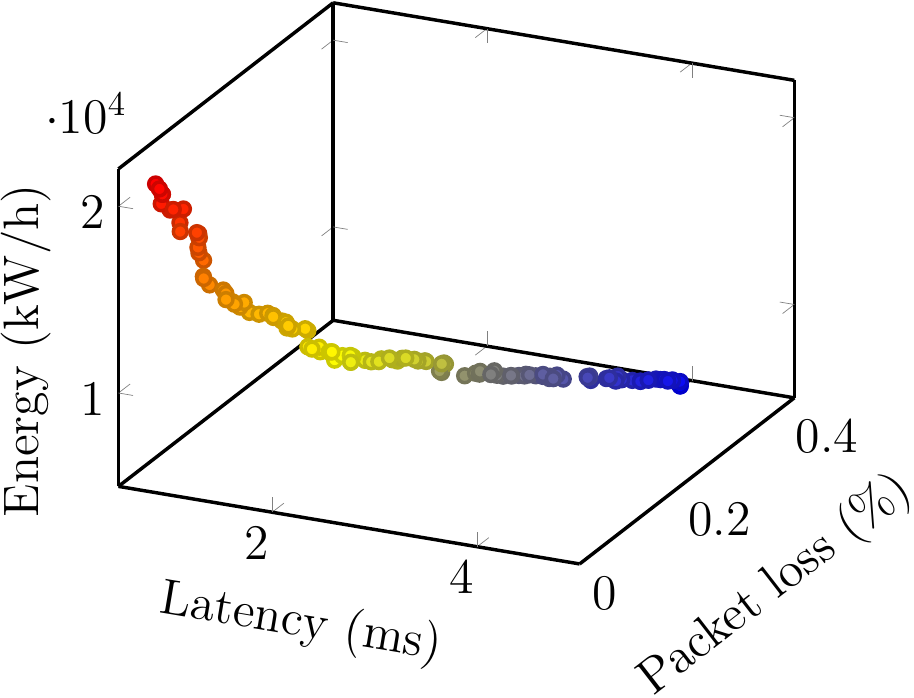}
        \caption{Proposed algorithm}
    \end{subfigure}
    \begin{subfigure}[b]{0.4\linewidth}
        \includegraphics[width=.8\textwidth]{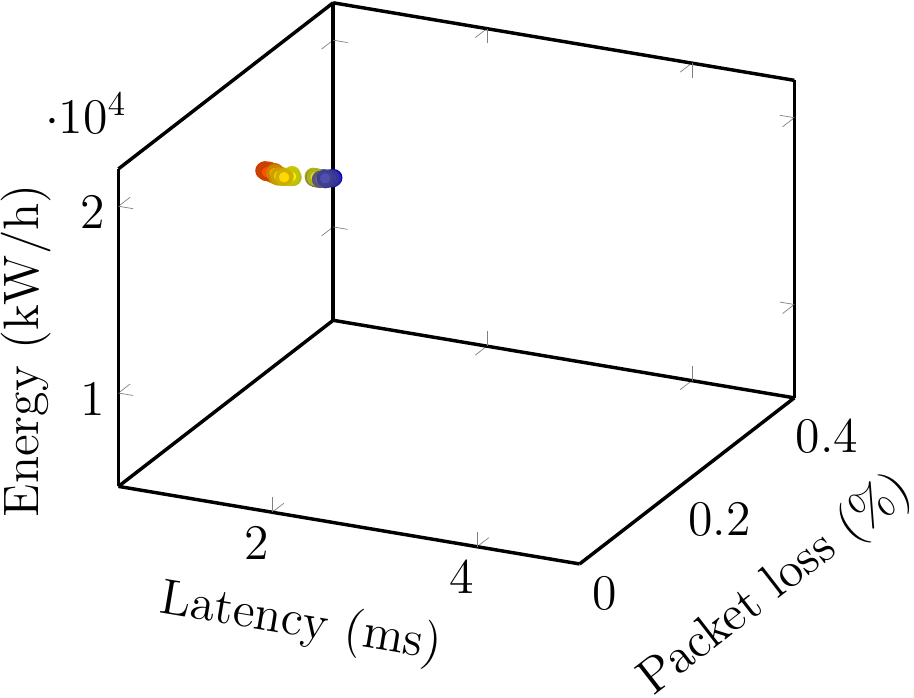}
        \caption{Direct representation}
    \end{subfigure}

    \vspace{1em}
    \caption{Non-dominated solutions obtained by NSGA-II using our proposed representation and the direct solution representation, respectively. The binary solution representation is omitted as it resulted in no feasible solution at all.}
    \label{fig:solution_representation_objectives}
\end{figure}

\subsubsection{Results}
From the results shown in Figs.~\ref{fig:solution_representation_comparison} and~\ref{fig:solution_representation_objectives}, we find that the solutions obtained by our proposed solution representation significantly outperforms alternative ones. Our proposed solution representation has two advantages over existing representations. First, our proposed representation guarantees feasible solutions irregardless of the input. In contrast, both the direct and binary solution representations were unable to find feasible solutions to larger problem instances. Second, our proposed representation integrates domain knowledge to generate solutions with shorter average distances between VNFs causing the resulting solutions to be closer to the Pareto front than alternative representations. Existing solution representations do not utilize this information and instead rely solely on the optimization framework to locate high quality solutions ineffectively.

Although the binary solution representation has been successfully applied to solve the VNFPP on small data centers (e.g.,~\cite{ChantreF20,KaurGK020,CharismiadisTPM20}), it does not scale well in the larger-scale problems considered in our experiments. With a binary solution representation, multiple VNFs to be placed on the same VM. This greatly complicates the search process compared to the direct or proposed solution representations with which this constraint is impossible to violate.

The direct solution representation is also only able to obtain feasible solutions to small data centers, as shown in~\pref{fig:solution_representation_objectives}, and exclusively finds solutions with high energy consumption. In particular, since a solution is only feasible when there is an instance of each VNF, solutions with more VNFs are more likely to be feasible than those with less VNFs and lower energy consumption. This leads the algorithm with the direct representation to be biased towards solutions with a high energy consumption. On larger problems with a large amount of VNFs, the direct solution representation is unable to find a solution with at least one instance of each VNF.

\begin{tcolorbox}[sharpish corners, top=2pt, bottom=2pt, left=4pt, right=4pt, boxrule=0.0pt, colback=black!5!white,leftrule=0.75mm,]
    \textbf{\underline{Response to RQ4:}} \textit{The proposed solution representation allows the algorithm to discover significantly higher quality solutions than alternative solution representations.}
\end{tcolorbox}
\vspace{-1.0em}

\subsection{Comparison with Other Approaches}
\label{sec:state_of_the_art}

\subsubsection{Methods}
To answer RQ5, we compare the performance of our tailored EMO algorithm with five SOTA peer algorithms for solving VNFPPs. Specifically, the following two meta-heuristic algorithms use the NSGA-II as the baseline but use different genetic operators.
\begin{itemize}
    \item\underline{Binary representation}: In \cite{ChantreF20}, a string of binary digits are used to represent the placement of primary and secondary VNFs. To implement a fair comparison, we only consider the placement of the primary VNFs.
    \item\underline{Previous work}: We also compare our algorithm against our earlier work on this topic \cite{BillingsleyLMMG19}. This work utilized a direct representation with a simple initialization strategy.
\end{itemize}
The three heuristic algorithms are as follows.
\begin{itemize}
    \item\underline{BFDSU}~\cite{ZhangXLLGW17}: This is a modified best-fit decreasing heuristic that considers each VNF in turn and selects a server that can accommodate the VNF according to a predefined probability. In addition, the result is weighted towards selecting a server with a lower capacity.
    \item\underline{ESP-VDCE}~\cite{KaurGK020}: This is specifically designed for the Fat Tree data centers. It uses a best fit approach but exclusively considers the servers nearest to where other VNFs of the same service have been placed.
    \item\underline{Stringer}~\cite{ChuaWZSH16}: This is also designed specifically for the Fat Tree data centers and it uses a round-robin placement strategy to place each VNF of each service in a sequence. The heuristic limits the available resources of each service and places a VNF on the first server with a sufficient capacity. If there is insufficient capacity in the data center for a VNF, the resources of each server are increased and the heuristic restarts from the first server.
\end{itemize}
Note that these heuristics assume that the number of service instances is known \textit{a priori}. Since each heuristic can only obtain a single solution, we generate a set of subproblems, each of which has a different number of service instances and is independently solved by a heuristic, to obtain a population of solutions at the end. In particular, we use the following two strategies to generate subproblems in our experiments.
\begin{itemize}
    \item One is to use the initialization operator developed in~\pref{sec:custom_operators} to serve our purpose. For the $i$th subproblem, the number of instances of the service $s$ is calculated as $N^I_{i,s}$ in~\pref{eq:num_services}.

    \item The other is to use the population obtained by our tailored EMO algorithm as a reference. For the $i$th subproblem, the number of instances of each service is the same as the $i$th solution obtained by our tailored EMO algorithm.
\end{itemize}
In our experiments, we generate $30$ VNFPP instances for six data centers with different sizes. To compare the performance of different algorithms, we use the QoS model developed in~\pref{sec:system_model} to evaluate the objective functions of the solutions obtained by different algorithms.

\subsubsection{Results}

\begin{figure*}[t!]
    \centering
    \includegraphics[width=\textwidth]{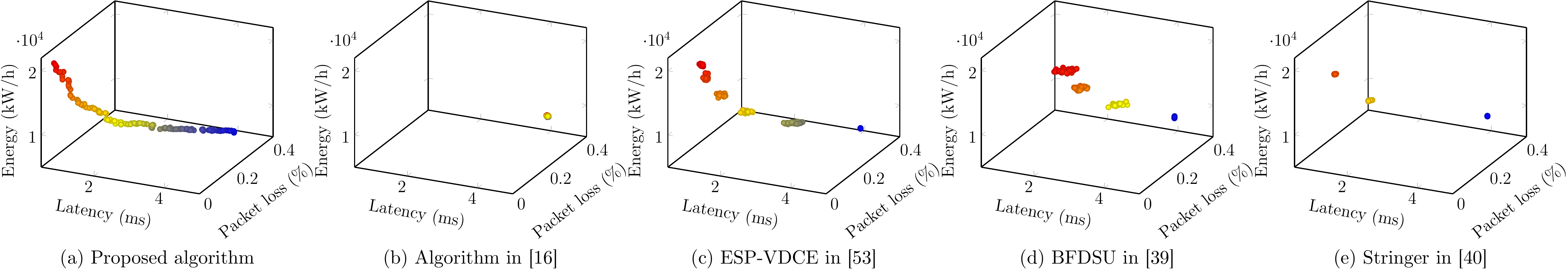}
    \caption{Non-dominated solutions obtained by NSGA-II using our proposed algorithm and algorithms from the literature. Subproblems for the heuristic were generated using our proposed initialization operator. The binary solution representation is omitted as it resulted in no feasible solution at all.}
    \label{fig:alg_objectives}
\end{figure*}

\begin{figure*}
    \centering
    \hfill
    \begin{minipage}[t]{.48\textwidth}
        \centering
        \includegraphics[width=\columnwidth]{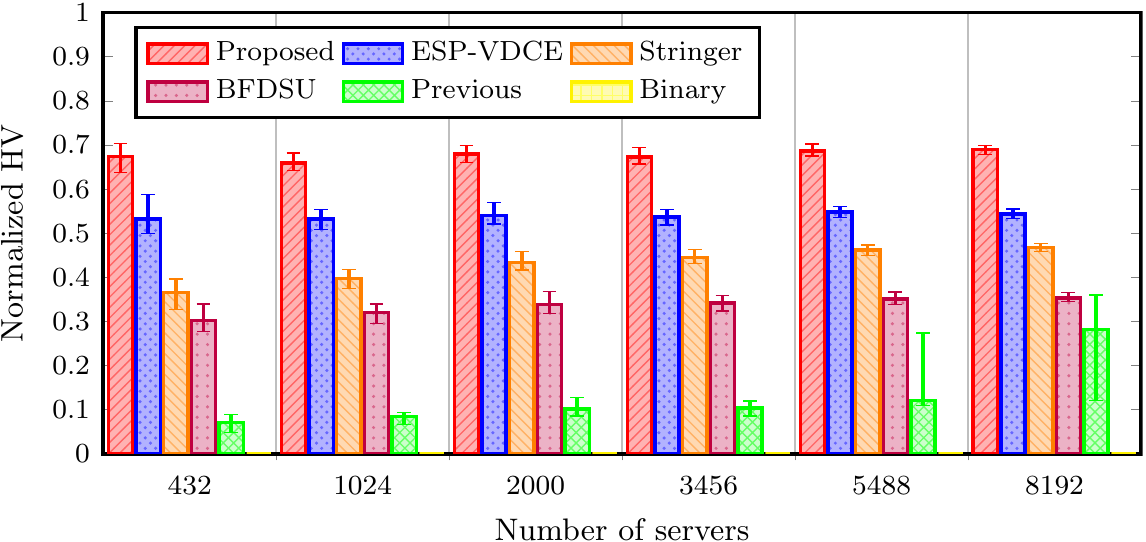}
        \caption{The lower, median, and upper quartiles of the HV values obtained by different algorithms on $30$ VNFPP instances using the initialization operator to generate subproblems for the heuristics.}
        \label{fig:alg_comparison}
    \end{minipage}\hfill
    \begin{minipage}[t]{.48\textwidth}
        \centering
        \includegraphics[width=\columnwidth]{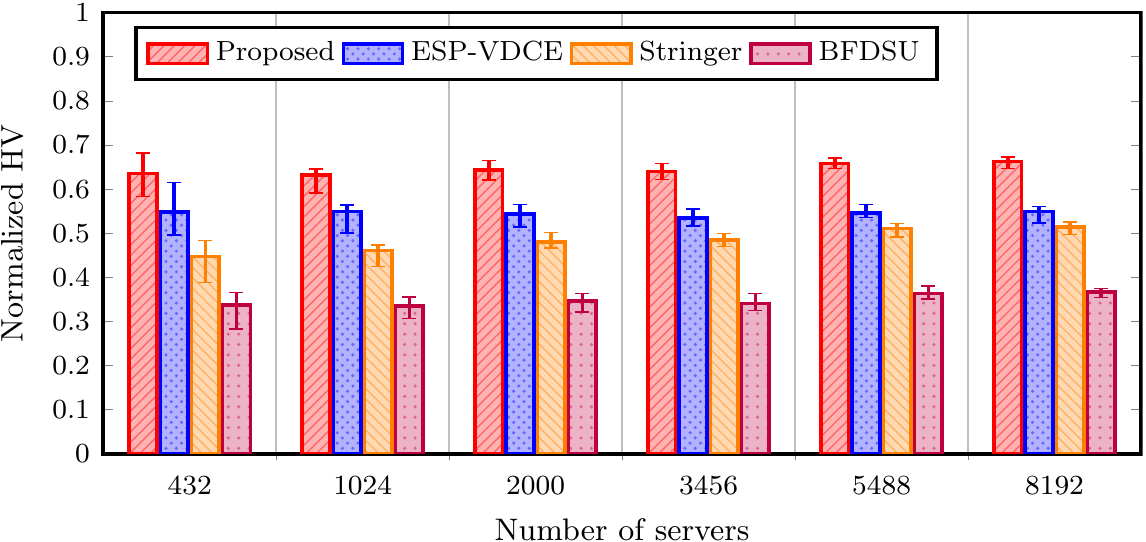}
        \caption{The lower, median, and upper quartiles of the HV values obtained by different algorithms on $30$ VNFPP instances using the solutions of our proposed algorithm to generate subproblems for the heuristics.}
        \label{fig:alg_fixed}
    \end{minipage}
    \hfill
\end{figure*}

From the results shown in~\pref{fig:alg_comparison} and~\pref{fig:alg_fixed}, it is clear that our proposed algorithm outperforms other competitors on all test cases. This can be attributed to our two proposed operators. First, it is clear from \pref{fig:alg_objectives} that proposed operators enable a diverse population of solutions. Our two proposed operators work together towards this goal. The initialization operator produces a diverse range of possible solutions, whilst the solution representation ensures that these solutions are feasible.

Second, our proposed solution representation minimizes the distance between sequential VNFs, improving the overall QoS. We note that the two best performing algorithms, our proposed algorithm and ESP-VDCE, aim to minimize the distance between sequential VNFs. In contrast, both BFDSU and Stringer tend to produce longer path lengths thus leading to significantly worse solutions than our proposed algorithm. Since Stringer restricts the capacity of each server, it causes services to be placed across multiple servers. Likewise, the stochastic component of BFDSU can cause it to place VNFs far away from any other VNF of the service. In contrast, our proposed algorithm incorporates useful information into the optimization process and places sequential VNFs close by thus leading to better solutions.

A final benefit of our algorithm is that it can iteratively improve the placements to minimize the energy consumption and QoS. Although ESP-VDCE does consider the path length, it otherwise uses a simple first fit heuristic that cannot consider how service instances should be placed in relation to each other. As a consequence, the performance of ESP-VDCE depends on the order in which services are considered. Our proposed algorithm considers the problem holistically and can make informed placement decisions.

\begin{tcolorbox}[sharpish corners, top=2pt, bottom=2pt, left=4pt, right=4pt, boxrule=0.0pt, colback=black!5!white,leftrule=0.75mm,]
    \textbf{\underline{Response to RQ5:}} \textit{Our proposed tailored EMO algorithm obtains significantly better solutions in terms of both convergence and diversity compared to other SOTA peer algorithms.}
\end{tcolorbox}
\vspace{-1.0em}

\subsection{Effectiveness on Constraint Handling}
\label{sec:custom_operators}

\subsubsection{Methods}
RQ6 aims to validate the effectiveness of our proposed solution representation for handling challenging anti-affinity and limited licenses constraints. We generated $30$ problem instances for a small data center with $412$ servers\footnote{Only the small data center is considered here in view of the poor scalability of the direct representation reported in~\pref{sec:state_of_the_art}.}. Note that we only compare our proposed algorithm with the meta-heuristic approach with the direct solution representation in our experiments given the poor performance of the binary solution representation reported in~\pref{sec:state_of_the_art} and the inability of the heuristic approaches to solve constrained VNFPPs.

For the anti-affinity constraints, we considered different numbers of anti-affinity services. Similarly, for the limited licenses constraints we considered different numbers of limited license VNFs and different numbers of licenses for each VNF. Since different VNFs have different service rates, we calculate the expected maximum number of instances of each VNF that could be accommodated by the data center (i.e. the maximum value of $N^I$ in the \pref{eq:num_services}) and restrict the solution to use a fraction of this number of licenses.

\subsubsection{Results}
\begin{figure}[t!]
    \centering
    \includegraphics[width=.6\linewidth]{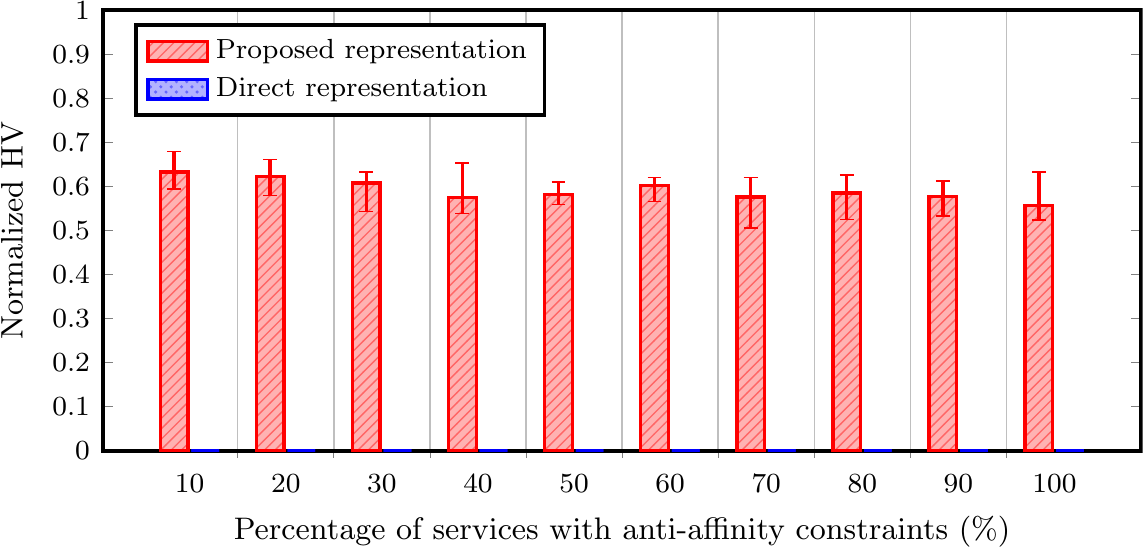}
    \caption{The lower quartile, median and upper quartile of the HV from our proposed solution representation and the direct solution representation. The direct solution representation resulted in no feasible solution at all.}
    \label{fig:anti_affinity}
\end{figure}

\begin{figure}[t!]
    \centering
    \begin{subfigure}[b]{0.48\linewidth}
        \includegraphics[width=\textwidth]{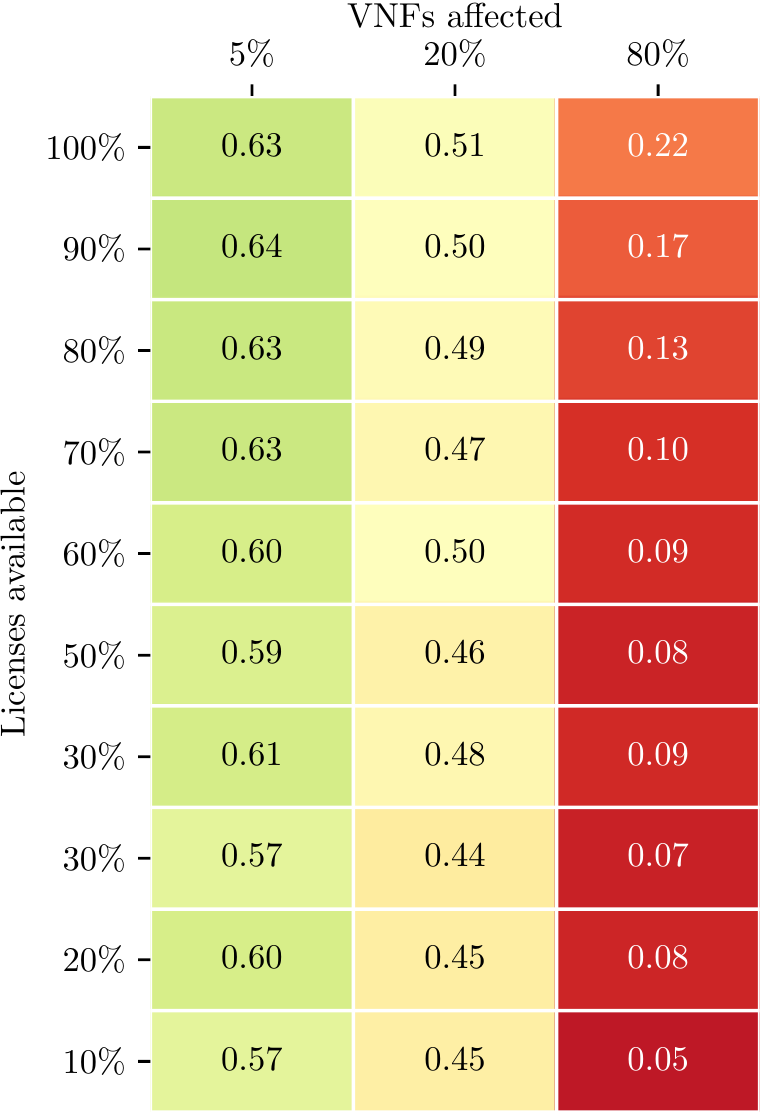}
        \caption{Proposed}
    \end{subfigure}
    \vspace{1em}
    \begin{subfigure}[b]{0.3732\linewidth}
        \includegraphics[width=\textwidth]{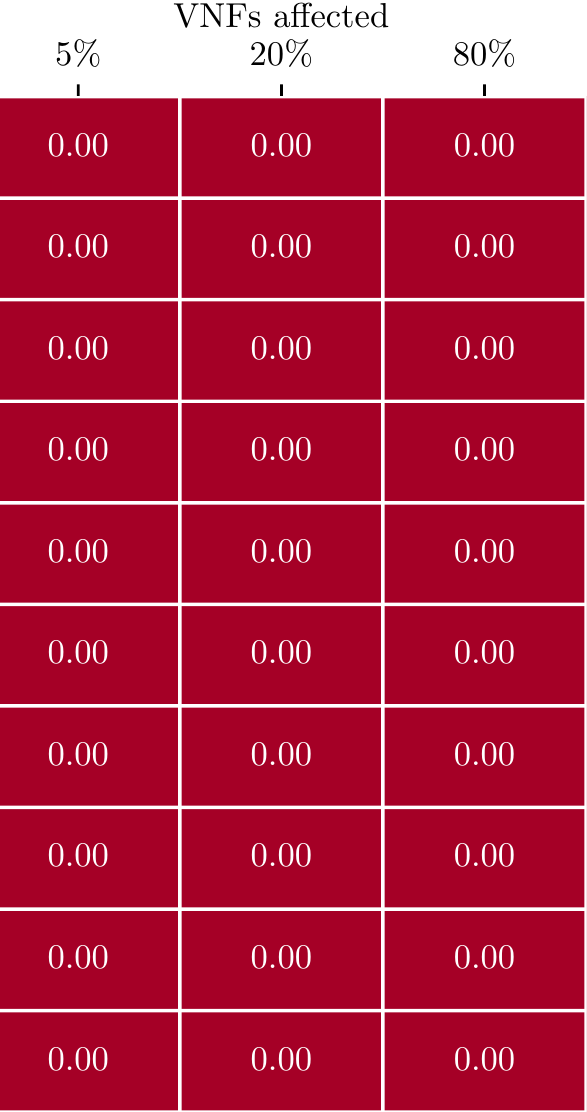}
        \caption{Direct representation}
    \end{subfigure}
    \vspace{1em}

    \hspace{2em}
    \includegraphics[width=.8\linewidth]{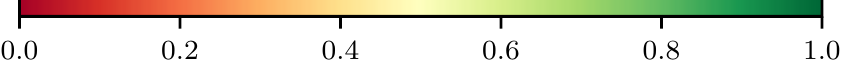}
    \caption{Mean normalized HV considering the proportion of VNFs ($10\%$--$90\%$) that have some restrictions and the number of licenses available ($80\%$, $20\%$, $5\%$) of the expected amount.}
    \label{fig:limited_licenses}
\end{figure}

From the results shown in Figs.~\ref{fig:anti_affinity} and \ref{fig:limited_licenses}, it is clear to see that the direct representation cannot find any feasible solution due to the narrow feasible search space. In contrast, our proposed operators are still able to find a diverse set of feasible solutions even for these highly constrained problems. As shown in \pref{fig:anti_affinity}, our algorithm produces consistently good results on the anti-affinity problems. This is a benefit of our proposed solution representation which ensures the satisfaction of the anti-affinity constraints. Since the solutions are guaranteed to be feasible, the algorithm should only optimize the number and placement of service instances. Any degradation in the HV indicator can be attributed to the narrower feasible search space causing better alternatives to be infeasible. In particular, anti-affinity constraints prevent VNFs of other services from being placed on a server thus can prevents a server from being fully utilized~\cite{LiWKC13,RuanLDL20,SunL20,LiNGY22}.

According to the results, the limited licenses constraints appear to be more challenging. As in~\pref{fig:limited_licenses}, populations obtained by our proposed algorithm have a better HV value when more licenses is allowed, whereas it falls down when fewer licenses are available. Furthermore, the percentage of VNFs that are affected has little impact. The lower HV values can be explained by a loss of diversity as a result of the feasible solution space being constrained. If any VNF in a service is constrained by a limited license constraint, this limits the number of service instances that can be placed. Hence the percentage of VNFs that can be placed is less significant as it is likely that a VNF in the service is already constrained.

\begin{tcolorbox}[sharpish corners, top=2pt, bottom=2pt, left=4pt, right=4pt, boxrule=0.0pt, colback=black!5!white,leftrule=0.75mm,]
    \textbf{\underline{Response to RQ6:}} \textit{From our empirical results, we find that our proposed genotype-phenotype solution representation is superior for handling highly constrained VNFPP.}
\end{tcolorbox}


%% file: conclusions.tex

\section{Conclusion}
\label{sec:conclusion}

By utilizing data center resources efficiently, we can provide high quality services and minimize their environmental impact. This work provided an efficient and accurate analytical model with which to evaluate the QoS of large data centers. To solve our VNFPP, we proposed a problem specific solution representation along with a tailored initialization strategy to guarantee the generation of feasible solutions, both of which are directly pluggable into any EMO algorithm. There are four main findings from our comprehensive experiments.

There are some disadvantages and extensions to our current approach that could be considered in future work.
\begin{itemize}
    \item Although this work only consider the Fat Tree network topology, our proposed heuristic---prefer to place VNFs on nearby servers---is applicable to any type of data center. Therefore, one interesting extension of this work is the consideration of arbitrary topologies.
    \item Although the execution time is not a priority in this work, it is notable that meta-heuristic approaches are typically slower than heuristic algorithms since they require a large number of model evaluations. That said, the significant improvements we make over existing heuristic alternatives justifies our approach. In future work, fast heuristic alternatives to accurate models may reduce this gap between heuristic and meta-heuristic algorithms.
    \item Last but not the least, it is interesting to investigate how an alternative problem formulation, with a different type of VNF and service, affect the design and results of a meta-heuristic alternative~\cite{WuLKZZ19,WuLKZZ17,LiC23,WilliamsLM23,LyuYWHL23}.
\end{itemize}

%% file: main.bbl
\begin{thebibliography}{100}
\providecommand{\url}[1]{#1}
\csname url@samestyle\endcsname
\providecommand{\newblock}{\relax}
\providecommand{\bibinfo}[2]{#2}
\providecommand{\BIBentrySTDinterwordspacing}{\spaceskip=0pt\relax}
\providecommand{\BIBentryALTinterwordstretchfactor}{4}
\providecommand{\BIBentryALTinterwordspacing}{\spaceskip=\fontdimen2\font plus
\BIBentryALTinterwordstretchfactor\fontdimen3\font minus
  \fontdimen4\font\relax}
\providecommand{\BIBforeignlanguage}[2]{{%
\expandafter\ifx\csname l@#1\endcsname\relax
\typeout{** WARNING: IEEEtran.bst: No hyphenation pattern has been}%
\typeout{** loaded for the language `#1'. Using the pattern for}%
\typeout{** the default language instead.}%
\else
\language=\csname l@#1\endcsname
\fi
#2}}
\providecommand{\BIBdecl}{\relax}
\BIBdecl

\bibitem{AndraeE15}
A.~Andrae and T.~Edler, ``On global electricity usage of communication
  technology: trends to 2030,'' \emph{Challenges}, vol.~6, no.~1, pp. 117--157,
  2015.

\bibitem{AvgerinouBC17}
M.~Avgerinou, P.~Bertoldi, and L.~Castellazzi, ``Trends in data centre energy
  consumption under the {European} code of conduct for data centre energy
  efficiency,'' \emph{Energies}, vol.~10, no. 1470, pp. 1--18, 2017.

\bibitem{DoddAGC20}
N.~Dodd, F.~Alfieri, M.~N. D. O.~G. Caldas, L.~M.-D.~J. Viegand, S.~Flucker,
  R.~Tozer, B.~Whitehead, and A.~W.~F. Brocklehurst, ``Development of the {EU}
  green public procurement {(GPP)} criteria for data centres,'' Server Rooms
  and Cloud Services, Publications Office of the European Union, Tech. Rep.,
  2020.

\bibitem{ShehabiARSSD16}
\BIBentryALTinterwordspacing
A.~Shehabi, S.~J. Smith, D.~A. Sartor, R.~E. Brown, M.~Herrlin, J.~G. Koomey,
  E.~R. Masanet, N.~Horner, I.~L. Azevedo, and W.~Lintner, ``{United States}
  data center energy usage report,'' Tech. Rep., 2016. [Online]. Available:
  \url{https://eta.lbl.gov/publications/united-states-data-center-energy}
\BIBentrySTDinterwordspacing

\bibitem{LuizelliCBG17}
M.~C. Luizelli, W.~L. da~Costa~Cordeiro, L.~S. Buriol, and L.~P. Gaspary, ``A
  fix-and-optimize approach for efficient and large scale virtual network
  function placement and chaining,'' \emph{Comput. Commun.}, vol. 102, pp.
  67--77, 2017.

\bibitem{SangJGDY17}
Y.~Sang, B.~Ji, G.~R. Gupta, X.~Du, and L.~Ye, ``Provably efficient algorithms
  for joint placement and allocation of virtual network functions,'' in
  \emph{{INFOCOM}'17: Proc. of the 2017 {IEEE} Conference on Computer
  Communications}, 2017, pp. 1--9.

\bibitem{CohenLNR15}
R.~Cohen, L.~Lewin{-}Eytan, J.~Naor, and D.~Raz, ``Near optimal placement of
  virtual network functions,'' in \emph{{INFOCOM}'15: Proc. of the 2015 {IEEE}
  Conference on Computer Communications}, 2015, pp. 1346--1354.

\bibitem{AddisBBS15}
B.~Addis, D.~Belabed, M.~Bouet, and S.~Secci, ``Virtual network functions
  placement and routing optimization,'' in \emph{CLOUDNET'15: Proc. of the 4th
  {IEEE} International Conference on Cloud Networking}, 2015, pp. 171--177.

\bibitem{JemaaPP16}
F.~B. Jemaa, G.~Pujolle, and M.~Pariente, ``Analytical models for qos-driven
  {VNF} placement and provisioning in wireless carrier cloud,'' in
  \emph{MSWiM'16: Proceedings of the 19th {ACM} International Conference on
  Modeling, Analysis and Simulation of Wireless and Mobile Systems}, 2016, pp.
  148--155.

\bibitem{GaoABS18}
M.~Gao, B.~Addis, M.~Bouet, and S.~Secci, ``Optimal orchestration of virtual
  network functions,'' \emph{Computer Networks}, vol. 142, pp. 108--127, 2018.

\bibitem{LakshmiI2013}
C.~Lakshmi and S.~A. Iyer, ``Application of queueing theory in health care: A
  literature review,'' \emph{Operations Research for Health Care}, vol.~2, no.
  1-2, pp. 25--39, 2013.

\bibitem{PapadopoulosC96}
H.~Papadopoulos and C.~Heavey, ``Queueing theory in manufacturing systems
  analysis and design: A classification of models for production and transfer
  lines,'' \emph{Eur. J. Oper. Res.}, vol.~92, no.~1, pp. 1--27, 1996.

\bibitem{OljiraGTB17}
D.~B. Oljira, K.~Grinnemo, J.~Taheri, and A.~Brunstr{\"{o}}m, ``A model for
  qos-aware {VNF} placement and provisioning,'' in \emph{{NFV-SDN}'17:
  Conference on Network Function Virtualization and Software Defined Networks},
  2017, pp. 1--7.

\bibitem{MarottaZDK17}
A.~Marotta, E.~Zola, F.~D'Andreagiovanni, and A.~Kassler, ``A fast robust
  optimization-based heuristic for the deployment of green virtual network
  functions,'' \emph{J. Network and Computer Applications}, vol.~95, pp.
  42--53, 2017.

\bibitem{LeivadeasFLIK18}
A.~Leivadeas, M.~Falkner, I.~Lambadaris, M.~Ibnkahla, and G.~Kesidis,
  ``Balancing delay and cost in virtual network function placement and
  chaining,'' in \emph{{NetSoft}'18: 4th {IEEE} Conference on Network
  Softwarization and Workshops}, 2018, pp. 433--440.

\bibitem{BillingsleyLMMG19}
J.~Billingsley, K.~Li, W.~Miao, G.~Min, and N.~Georgalas, ``A formal model for
  multi-objective optimisation of network function virtualisation placement,''
  in \emph{EMO'19: Proc. of 10th International Conference on Evolutionary
  Multi-Criterion Optimization}, ser. Lecture Notes in Computer Science, vol.
  11411.\hskip 1em plus 0.5em minus 0.4em\relax Springer, 2019, pp. 529--540.

\bibitem{BariCAB15}
M.~F. Bari, S.~R. Chowdhury, R.~Ahmed, and R.~Boutaba, ``On orchestrating
  virtual network functions,'' in \emph{CNSM'11: Proc. of the 11th
  International Conference on Network and Service Management}, 2015, pp.
  50--56.

\bibitem{KawashimaOOM16}
K.~Kawashima, T.~Otoshi, Y.~Ohsita, and M.~Murata, ``Dynamic placement of
  virtual network functions based on model predictive control,'' in
  \emph{{NOMS}'16: Network Operations and Management Symposium}, 2016, pp.
  1037--1042.

\bibitem{AllegKMA17}
A.~Alleg, R.~Kouah, S.~Moussaoui, and T.~Ahmed, ``Virtual network functions
  placement and chaining for real-time applications,'' in \emph{CAMAD'17: Proc.
  of the 22nd {IEEE} International Workshop on Computer Aided Modeling and
  Design of Communication Links and Networks}, 2017, pp. 1--6.

\bibitem{WangJ20}
H.~Wang and Y.~Jin, ``A random forest-assisted evolutionary algorithm for
  data-driven constrained multiobjective combinatorial optimization of trauma
  systems,'' \emph{{IEEE} Trans. Cybern.}, vol.~50, no.~2, pp. 536--549, 2020.

\bibitem{ZhouL17}
M.~Zhou and J.~Liu, ``A two-phase multiobjective evolutionary algorithm for
  enhancing the robustness of scale-free networks against multiple malicious
  attacks,'' \emph{{IEEE} Trans. Cybern.}, vol.~47, no.~2, pp. 539--552, 2017.

\bibitem{WangWZ19}
J.~Wang, T.~Weng, and Q.~Zhang, ``A two-stage multiobjective evolutionary
  algorithm for multiobjective multidepot vehicle routing problem with time
  windows,'' \emph{{IEEE} Trans. Cybern.}, vol.~49, no.~7, pp. 2467--2478,
  2019.

\bibitem{LiuLJ14}
C.~Liu, J.~Liu, and Z.~Jiang, ``A multiobjective evolutionary algorithm based
  on similarity for community detection from signed social networks,''
  \emph{{IEEE} Trans. Cybern.}, vol.~44, no.~12, pp. 2274--2287, 2014.

\bibitem{LiDZK15}
K.~Li, K.~Deb, Q.~Zhang, and S.~Kwong, ``An evolutionary many-objective
  optimization algorithm based on dominance and decomposition,'' \emph{{IEEE}
  Trans. Evol. Comput.}, vol.~19, no.~5, pp. 694--716, 2015.

\bibitem{LiCFY19}
K.~Li, R.~Chen, G.~Fu, and X.~Yao, ``Two-archive evolutionary algorithm for
  constrained multiobjective optimization,'' \emph{{IEEE} Trans. Evol.
  Comput.}, vol.~23, no.~2, pp. 303--315, 2019.

\bibitem{LiKZD15}
K.~Li, S.~Kwong, Q.~Zhang, and K.~Deb, ``Interrelationship-based selection for
  decomposition multiobjective optimization,'' \emph{{IEEE} Trans. Cybern.},
  vol.~45, no.~10, pp. 2076--2088, 2015.

\bibitem{LiZKLW14}
K.~Li, Q.~Zhang, S.~Kwong, M.~Li, and R.~Wang, ``Stable matching-based
  selection in evolutionary multiobjective optimization,'' \emph{{IEEE} Trans.
  Evol. Comput.}, vol.~18, no.~6, pp. 909--923, 2014.

\bibitem{LiFKZ14}
K.~Li, {\'{A}}.~Fialho, S.~Kwong, and Q.~Zhang, ``Adaptive operator selection
  with bandits for a multiobjective evolutionary algorithm based on
  decomposition,'' \emph{{IEEE} Trans. Evol. Comput.}, vol.~18, no.~1, pp.
  114--130, 2014.

\bibitem{CaoZACHS16}
J.~Cao, Y.~Zhang, W.~An, X.~Chen, Y.~Han, and J.~Sun, ``{VNF} placement in
  hybrid {NFV} environment: Modeling and genetic algorithms,'' in
  \emph{ICPADS'16: Proc. of the 22nd {IEEE} International Conference on
  Parallel and Distributed Systems}, 2016, pp. 769--777.

\bibitem{RankothgeLRL17}
W.~Rankothge, F.~Le, A.~Russo, and J.~Lobo, ``Optimizing resource allocation
  for virtualized network functions in a cloud center using genetic
  algorithms,'' \emph{{IEEE} Trans. Network and Service Management}, vol.~14,
  no.~2, pp. 343--356, 2017.

\bibitem{LangeGZTJ17}
S.~Lange, A.~Grigorjew, T.~Zinner, P.~Tran{-}Gia, and M.~Jarschel, ``A
  multi-objective heuristic for the optimization of virtual network function
  chain placement,'' in \emph{{ITC}'17: 29th International Teletraffic
  Congress}, 2017, pp. 152--160.

\bibitem{BillingsleyLMMG20}
J.~Billingsley, K.~Li, W.~Miao, G.~Min, and N.~Georgalas, ``Routing-led
  placement of vnfs in arbitrary networks,'' in \emph{CEC:20: Proc. of 2020
  {IEEE} Congress on Evolutionary Computation}.\hskip 1em plus 0.5em minus
  0.4em\relax {IEEE}, 2020, pp. 1--8.

\bibitem{BillingsleyMLMG20}
J.~Billingsley, W.~Miao, K.~Li, G.~Min, and N.~Georgalas, ``Performance
  analysis of {SDN} and {NFV} enabled mobile cloud computing,'' in
  \emph{GLOBECOM'20: Proc. of 2020 {IEEE} Global Communications
  Conference}.\hskip 1em plus 0.5em minus 0.4em\relax {IEEE}, 2020, pp. 1--6.

\bibitem{BillingsleyLMMG21}
J.~Billingsley, K.~Li, W.~Miao, G.~Min, and N.~Georgalas, ``Parallel algorithms
  for the multiobjective virtual network function placement problem,'' in
  \emph{EMO'21: Proc. of 11th International Conference on Evolutionary
  Multi-Criterion Optimization}, ser. Lecture Notes in Computer Science, vol.
  12654.\hskip 1em plus 0.5em minus 0.4em\relax Springer, 2021, pp. 708--720.

\bibitem{Landa-Silva13}
D.~Landa{-}Silva, ``Franz rothlauf: Design of modern heuristics - springer,
  2011, {ISBN} 978-3-540-72961-7,'' \emph{Genet. Program. Evolvable Mach.},
  vol.~14, no.~1, pp. 119--121, 2013.

\bibitem{IntelDPDK}
``Impact of the intel data plane development kit (intel dpdk) on packet
  throughput in virtualized network elements,'' Intel, Tech. Rep., 2013.

\bibitem{IntelPPP}
``Packet processing performance of virtualized platforms with linux* and intel
  architecture,'' Intel, Tech. Rep., 2013.

\bibitem{MiottoLCG19}
G.~Miotto, M.~C. Luizelli, W.~L. da~Costa~Cordeiro, and L.~P. Gaspary,
  ``Adaptive placement {\&} chaining of virtual network functions with
  {NFV-PEAR},'' \emph{J. Internet Services and Applications}, vol.~10, no.~1,
  pp. 3:1--3:19, 2019.

\bibitem{BaumgartnerRB15}
A.~Baumgartner, V.~S. Reddy, and T.~Bauschert, ``Combined virtual mobile core
  network function placement and topology optimization with latency bounds,''
  in \emph{EWSDN'15: Proc. of the 4th European Workshop on Software Defined
  Networks}, 2015, pp. 97--102.

\bibitem{GuoWLQA0Y20}
H.~Guo, Y.~Wang, Z.~Li, X.~Qiu, H.~An, P.~Yu, and N.~Yuan, ``Cost-aware
  placement and chaining of service function chain with {VNF} instance
  sharing,'' in \emph{{NOMS}'20: Proc. of the 2020 {IEEE/IFIP} Network
  Operations and Management Symposium}.\hskip 1em plus 0.5em minus 0.4em\relax
  {IEEE}, 2020, pp. 1--8.

\bibitem{Katz53}
L.~Katz, ``A new status index derived from sociometric analysis,''
  \emph{Psychometrika}, vol.~18, no.~1, pp. 39--43, 1953.

\bibitem{QiSW19}
D.~Qi, S.~Shen, and G.~Wang, ``Towards an efficient {VNF} placement in network
  function virtualization,'' \emph{Comput. Commun.}, vol. 138, pp. 81--89,
  2019.

\bibitem{QuASK17}
L.~Qu, C.~Assi, K.~B. Shaban, and M.~J. Khabbaz, ``A reliability-aware network
  service chain provisioning with delay guarantees in nfv-enabled enterprise
  datacenter networks,'' \emph{IEEE Trans. Netw. Serv. Manag.}, vol.~14, no.~3,
  pp. 554--568, 2017.

\bibitem{HawiloJS19}
H.~Hawilo, M.~Jammal, and A.~Shami, ``Network function virtualization-aware
  orchestrator for service function chaining placement in the cloud,''
  \emph{{IEEE} J. Sel. Areas Commun.}, vol.~37, no.~3, pp. 643--655, 2019.

\bibitem{VizarretaCMMK17}
P.~Vizarreta, M.~Condoluci, C.~M. Machuca, T.~Mahmoodi, and W.~Kellerer,
  ``{QoS}-driven function placement reducing expenditures in {NFV}
  deployments,'' in \emph{{ICC}'17: Proc. of the 2017 {IEEE} International
  Conference on Communications}.\hskip 1em plus 0.5em minus 0.4em\relax {IEEE},
  2017, pp. 1--7.

\bibitem{BeckB15}
M.~T. Beck and J.~F. Botero, ``Coordinated allocation of service function
  chains,'' in \emph{{GLOBECOM}'15: Proc. of the IEEE 2015 Global
  Communications Conference}, 2015, pp. 1--6.

\bibitem{ZhangXLLGW17}
Q.~Zhang, Y.~Xiao, F.~Liu, J.~C.~S. Lui, J.~Guo, and T.~Wang, ``Joint
  optimization of chain placement and request scheduling for network function
  virtualization,'' in \emph{{ICDCS}'17: Proc. of the 2017 {IEEE} International
  Conference on Distributed Computing Systems}, K.~Lee and L.~Liu, Eds.\hskip
  1em plus 0.5em minus 0.4em\relax {IEEE} Computer Society, 2017, pp. 731--741.

\bibitem{ChuaWZSH16}
F.~C.~T. Chua, J.~Ward, Y.~Zhang, P.~Sharma, and B.~A. Huberman, ``Stringer:
  Balancing latency and resource usage in service function chain
  provisioning,'' \emph{{IEEE} Internet Comput.}, vol.~20, no.~6, pp. 22--31,
  2016.

\bibitem{GouarebFA18}
R.~Gouareb, V.~Friderikos, and A.~Aghvami, ``Virtual network functions routing
  and placement for edge cloud latency minimization,'' \emph{{IEEE} J. Sel.
  Areas Commun.}, vol.~36, no.~10, pp. 2346--2357, 2018.

\bibitem{AgarwalMCD18}
S.~Agarwal, F.~Malandrino, C.~Chiasserini, and S.~De, ``Joint {VNF} placement
  and {CPU} allocation in 5g,'' in \emph{{INFOCOM}'18: {IEEE} Conference on
  Computer Communications}.\hskip 1em plus 0.5em minus 0.4em\relax {IEEE},
  2018, pp. 1943--1951.

\bibitem{XueZB13}
B.~Xue, M.~Zhang, and W.~N. Browne, ``Particle swarm optimization for feature
  selection in classification: {A} multi-objective approach,'' \emph{{IEEE}
  Trans. Cybern.}, vol.~43, no.~6, pp. 1656--1671, 2013.

\bibitem{MavrovouniotisM17}
M.~Mavrovouniotis, F.~M. M{\"{u}}ller, and S.~Yang, ``Ant colony optimization
  with local search for dynamic traveling salesman problems,'' \emph{{IEEE}
  Trans. Cybern.}, vol.~47, no.~7, pp. 1743--1756, 2017.

\bibitem{YuanBTZLL17}
H.~Yuan, J.~Bi, W.~Tan, M.~Zhou, B.~H. Li, and J.~Li, ``{TTSA:} an effective
  scheduling approach for delay bounded tasks in hybrid clouds,'' \emph{{IEEE}
  Trans. Cybern.}, vol.~47, no.~11, pp. 3658--3668, 2017.

\bibitem{ChenZLGGZYCLZ19}
Z.~Chen, Z.~Zhan, Y.~Lin, Y.~Gong, T.~Gu, F.~Zhao, H.~Yuan, X.~Chen, Q.~Li, and
  J.~Zhang, ``Multiobjective cloud workflow scheduling: {A} multiple
  populations ant colony system approach,'' \emph{{IEEE} Trans. Cybern.},
  vol.~49, no.~8, pp. 2912--2926, 2019.

\bibitem{YoonK13}
Y.~Yoon and Y.~Kim, ``An efficient genetic algorithm for maximum coverage
  deployment in wireless sensor networks,'' \emph{{IEEE} Trans. Cybern.},
  vol.~43, no.~5, pp. 1473--1483, 2013.

\bibitem{JiaMZ21}
Y.-H. Jia, Y.~Mei, and M.~Zhang, ``A two-stage swarm optimizer with local
  search for water distribution network optimization,'' \emph{{IEEE} Trans.
  Cybern.}, 2021.

\bibitem{PengJW19}
X.~Peng, Y.~Jin, and H.~Wang, ``Multimodal optimization enhanced cooperative
  coevolution for large-scale optimization,'' \emph{{IEEE} Trans. Cybern.},
  vol.~49, no.~9, pp. 3507--3520, 2019.

\bibitem{ChengJ15}
R.~Cheng and Y.~Jin, ``A competitive swarm optimizer for large scale
  optimization,'' \emph{{IEEE} Trans. Cybern.}, vol.~45, no.~2, pp. 191--204,
  2015.

\bibitem{RankothgeMLRL15}
W.~Rankothge, J.~Ma, F.~Le, A.~Russo, and J.~Lobo, ``Towards making network
  function virtualization a cloud computing service,'' in \emph{IM'15: Proc. of
  the 2015 {IFIP/IEEE} International Symposium on Integrated Network
  Management}, 2015, pp. 89--97.

\bibitem{ChantreF20}
H.~D. Chantre and N.~L.~S. da~Fonseca, ``The location problem for the
  provisioning of protected slices in {NFV}-based {MEC} infrastructure,''
  \emph{{IEEE} J. Sel. Areas Commun.}, vol.~38, no.~7, pp. 1505--1514, 2020.

\bibitem{KaurGK020}
K.~Kaur, S.~Garg, G.~Kaddoum, and S.~Guo, ``{ESP-VDCE:} energy, sla, and
  price-driven virtual data center embedding,'' in \emph{{ICC}'20: Proc. of the
  IEEE 2020 International Conference on Communications}, 2020, pp. 1--7.

\bibitem{DebAPM02}
K.~Deb, S.~Agrawal, A.~Pratap, and T.~Meyarivan, ``A fast and elitist
  multiobjective genetic algorithm: {NSGA-II},'' \emph{{IEEE} Trans.
  Evolutionary Computation}, vol.~6, no.~2, pp. 182--197, 2002.

\bibitem{SoualahMGZ17}
O.~Soualah, M.~Mechtri, C.~Ghribi, and D.~Zeghlache, ``Energy efficient
  algorithm for {VNF} placement and chaining,'' in
  \emph{{CCGRID}'17:International Symposium on Cluster, Cloud and Grid
  Computing}.\hskip 1em plus 0.5em minus 0.4em\relax {IEEE} Computer Society /
  {ACM}, 2017, pp. 579--588.

\bibitem{ChenLY18}
R.~Chen, K.~Li, and X.~Yao, ``Dynamic multiobjectives optimization with a
  changing number of objectives,'' \emph{{IEEE} Trans. Evol. Comput.}, vol.~22,
  no.~1, pp. 157--171, 2018.

\bibitem{ZouJYZZL19}
J.~Zou, C.~Ji, S.~Yang, Y.~Zhang, J.~Zheng, and K.~Li, ``A knee-point-based
  evolutionary algorithm using weighted subpopulation for many-objective
  optimization,'' \emph{Swarm and Evolutionary Computation}, vol.~47, pp.
  33--43, 2019.

\bibitem{LiZZL09}
K.~Li, J.~Zheng, C.~Zhou, and H.~Lv, ``An improved differential evolution for
  multi-objective optimization,'' in \emph{CSIE'09: Proc. of 2009 {WRI} World
  Congress on Computer Science and Information Engineering}, 2009, pp.
  825--830.

\bibitem{LiZLZL09}
K.~Li, J.~Zheng, M.~Li, C.~Zhou, and H.~Lv, ``A novel algorithm for
  non-dominated hypervolume-based multiobjective optimization,'' in
  \emph{SMC'09: Proc. of 2009 the {IEEE} International Conference on Systems,
  Man and Cybernetics}, 2009, pp. 5220--5226.

\bibitem{Li19}
K.~Li, ``Progressive preference learning: Proof-of-principle results in
  {MOEA/D},'' in \emph{EMO'19: Proc. of the 10th International Conference
  Evolutionary Multi-Criterion Optimization}, 2019, pp. 631--643.

\bibitem{LiK14}
K.~Li and S.~Kwong, ``A general framework for evolutionary multiobjective
  optimization via manifold learning,'' \emph{Neurocomputing}, vol. 146, pp.
  65--74, 2014.

\bibitem{LiFK11}
K.~Li, {\'{A}}.~Fialho, and S.~Kwong, ``Multi-objective differential evolution
  with adaptive control of parameters and operators,'' in \emph{LION5: Proc. of
  the 5th International Conference on Learning and Intelligent Optimization},
  2011, pp. 473--487.

\bibitem{LiKWTM13}
K.~Li, S.~Kwong, R.~Wang, K.~Tang, and K.~Man, ``Learning paradigm based on
  jumping genes: {A} general framework for enhancing exploration in
  evolutionary multiobjective optimization,'' \emph{Inf. Sci.}, vol. 226, pp.
  1--22, 2013.

\bibitem{CaoKWL12}
J.~Cao, S.~Kwong, R.~Wang, and K.~Li, ``A weighted voting method using minimum
  square error based on extreme learning machine,'' in \emph{ICMLC'12: Proc. of
  the 2012 International Conference on Machine Learning and Cybernetics}, 2012,
  pp. 411--414.

\bibitem{CaoKWL14}
------, ``{AN} indicator-based selection multi-objective evolutionary algorithm
  with preference for multi-class ensemble,'' in \emph{ICMLC'14: Proc. of the
  2014 International Conference on Machine Learning and Cybernetics}, 2014, pp.
  147--152.

\bibitem{LiDZZ17}
K.~Li, K.~Deb, Q.~Zhang, and Q.~Zhang, ``Efficient nondomination level update
  method for steady-state evolutionary multiobjective optimization,''
  \emph{{IEEE} Trans. Cybernetics}, vol.~47, no.~9, pp. 2838--2849, 2017.

\bibitem{LiKD15}
K.~Li, S.~Kwong, and K.~Deb, ``A dual-population paradigm for evolutionary
  multiobjective optimization,'' \emph{Inf. Sci.}, vol. 309, pp. 50--72, 2015.

\bibitem{PoissonTraffic}
W.~{Miao}, G.~{Min}, Y.~{Wu}, H.~{Huang}, Z.~{Zhao}, H.~{Wang}, and C.~{Luo},
  ``Stochastic performance analysis of network function virtualization in
  future internet,'' \emph{{IEEE} J. Sel. Areas Commun.}, vol.~37, no.~3, pp.
  613--626, 2019.

\bibitem{InfiniteQueue}
R.~{Bolla}, R.~{Bruschi}, F.~{Davoli}, and J.~F. {Pajo}, ``A model-based
  approach towards real-time analytics in nfv infrastructures,'' \emph{IEEE
  Trans. Green Commun.}, vol.~4, no.~2, pp. 529--541, 2020.

\bibitem{Kleinrock75}
L.~Kleinrock, \emph{Theory, Volume 1, Queueing Systems}.\hskip 1em plus 0.5em
  minus 0.4em\relax Wiley-Interscience, 1975.

\bibitem{QuZYSLR20}
K.~Qu, W.~Zhuang, Q.~Ye, X.~Shen, X.~Li, and J.~Rao, ``Dynamic flow migration
  for embedded services in sdn/nfv-enabled 5g core networks,'' \emph{IEEE Trans
  Commun.}, vol.~68, no.~4, pp. 2394--2408, 2020.

\bibitem{AlFaresLV08}
M.~Al{-}Fares, A.~Loukissas, and A.~Vahdat, ``A scalable, commodity data center
  network architecture,'' in \emph{{SIGCOMM}'08 Conference on Applications,
  Technologies, Architectures, and Protocols for Computer Communications},
  2008, pp. 63--74.

\bibitem{KellerTLB12}
G.~Keller, M.~Tighe, H.~Lutfiyya, and M.~Bauer, ``An analysis of first fit
  heuristics for the virtual machine relocation problem,'' in \emph{{CNSM}'12:
  Proc. of the 8th International Conference on Network and Service Management},
  2012, pp. 406--413.

\bibitem{Hopps2000}
C.~Hopps, ``Rfc 2992: Analysis of an equal-cost multi-path algorithm,'' Tech.
  Rep., 2000.

\bibitem{ChiesaKS17}
M.~Chiesa, G.~Kindler, and M.~Schapira, ``Traffic engineering with
  equal-cost-multipath: An algorithmic perspective,'' \emph{{IEEE/ACM} Trans.
  Netw.}, vol.~25, no.~2, pp. 779--792, 2017.

\bibitem{LiKWCR12}
K.~Li, S.~Kwong, R.~Wang, J.~Cao, and I.~J. Rudas, ``Multi-objective
  differential evolution with self-navigation,'' in \emph{SMC'12: Proc. of the
  2012 {IEEE} International Conference on Systems, Man, and Cybernetics}, 2012,
  pp. 508--513.

\bibitem{LiWKC13}
K.~Li, R.~Wang, S.~Kwong, and J.~Cao, ``Evolving extreme learning machine
  paradigm with adaptive operator selection and parameter control,''
  \emph{International Journal of Uncertainty, Fuzziness and Knowledge-Based
  Systems}, vol. supp02, pp. 143--154, 2013.

\bibitem{CaoKWLLK15}
J.~Cao, S.~Kwong, R.~Wang, X.~Li, K.~Li, and X.~Kong, ``Class-specific soft
  voting based multiple extreme learning machines ensemble,''
  \emph{Neurocomputing}, vol. 149, pp. 275--284, 2015.

\bibitem{LiDY18}
K.~Li, K.~Deb, and X.~Yao, ``R-metric: Evaluating the performance of
  preference-based evolutionary multiobjective optimization using reference
  points,'' \emph{{IEEE} Trans. Evolutionary Computation}, vol.~22, no.~6, pp.
  821--835, 2018.

\bibitem{WuKZLWL15}
M.~Wu, S.~Kwong, Q.~Zhang, K.~Li, R.~Wang, and B.~Liu, ``Two-level stable
  matching-based selection in {MOEA/D},'' in \emph{SMC'15: Proc. of the 2015
  {IEEE} International Conference on Systems, Man, and Cybernetics}, 2015, pp.
  1720--1725.

\bibitem{LiKCLZS12}
K.~Li, S.~Kwong, J.~Cao, M.~Li, J.~Zheng, and R.~Shen, ``Achieving balance
  between proximity and diversity in multi-objective evolutionary algorithm,''
  \emph{Inf. Sci.}, vol. 182, no.~1, pp. 220--242, 2012.

\bibitem{LiDAY17}
K.~Li, K.~Deb, O.~T. Altin{\"{o}}z, and X.~Yao, ``Empirical investigations of
  reference point based methods when facing a massively large number of
  objectives: First results,'' in \emph{EMO'17: Proc. of the 9th International
  Conference Evolutionary Multi-Criterion Optimization}, 2017, pp. 390--405.

\bibitem{LiDZ15}
K.~Li, K.~Deb, and Q.~Zhang, ``Evolutionary multiobjective optimization with
  hybrid selection principles,'' in \emph{CEC'15: Proc. of 2015 {IEEE} Congress
  on Evolutionary Computation}.\hskip 1em plus 0.5em minus 0.4em\relax {IEEE},
  2015, pp. 900--907.

\bibitem{LiXT19}
K.~Li, Z.~Xiang, and K.~C. Tan, ``Which surrogate works for empirical
  performance modelling? {A} case study with differential evolution,'' in
  \emph{CEC'19: Proc. of the 2019 {IEEE} Congress on Evolutionary Computation},
  2019, pp. 1988--1995.

\bibitem{GaoNL19}
H.~Gao, H.~Nie, and K.~Li, ``Visualisation of pareto front approximation: {A}
  short survey and empirical comparisons,'' in \emph{CEC'19: Proc. of the 2019
  {IEEE} Congress on Evolutionary Computation}, 2019, pp. 1750--1757.

\bibitem{LiuLC19}
M.~Liu, K.~Li, and T.~Chen, ``Security testing of web applications: a
  search-based approach for detecting {SQL} injection vulnerabilities,'' in
  \emph{GECCO'19: Proc. of the 2019 Genetic and Evolutionary Computation
  Conference Companion}, 2019, pp. 417--418.

\bibitem{LiZ19}
K.~Li and Q.~Zhang, ``Decomposition multi-objective optimisation: current
  developments and future opportunities,'' in \emph{GECCO'19: Proc. of the 2019
  Genetic and Evolutionary Computation Conference Companion}, 2019, pp.
  1002--1031.

\bibitem{KumarBCLB18}
S.~Kumar, R.~Bahsoon, T.~Chen, K.~Li, and R.~Buyya, ``Multi-tenant cloud
  service composition using evolutionary optimization,'' in \emph{ICPADS'18:
  Proc. of the 24th {IEEE} International Conference on Parallel and Distributed
  Systems}, 2018, pp. 972--979.

\bibitem{CaoWKL11}
J.~Cao, H.~Wang, S.~Kwong, and K.~Li, ``Combining interpretable fuzzy
  rule-based classifiers via multi-objective hierarchical evolutionary
  algorithm,'' in \emph{SMC'11: Proc. of the 2011 {IEEE} International
  Conference on Systems, Man and Cybernetics}.\hskip 1em plus 0.5em minus
  0.4em\relax {IEEE}, 2011, pp. 1771--1776.

\bibitem{LiX0WT20}
K.~Li, Z.~Xiang, T.~Chen, S.~Wang, and K.~C. Tan, ``Understanding the automated
  parameter optimization on transfer learning for cross-project defect
  prediction: an empirical study,'' in \emph{{ICSE}'20: Proc. of the 42nd
  International Conference on Software Engineering}.\hskip 1em plus 0.5em minus
  0.4em\relax {ACM}, 2020, pp. 566--577.

\bibitem{LiuLC20}
M.~Liu, K.~Li, and T.~Chen, ``{DeepSQLi}: deep semantic learning for testing
  {SQL} injection,'' in \emph{{ISSTA}'20: Proc. of the 29th {ACM} {SIGSOFT}
  International Symposium on Software Testing and Analysis}.\hskip 1em plus
  0.5em minus 0.4em\relax {ACM}, 2020, pp. 286--297.

\bibitem{LiXCT20}
K.~Li, Z.~Xiang, T.~Chen, and K.~C. Tan, ``{BiLO-CPDP}: Bi-level programming
  for automated model discovery in cross-project defect prediction,'' in
  \emph{ASE'20: Proc. of the 35th {IEEE/ACM} International Conference on
  Automated Software Engineering}.\hskip 1em plus 0.5em minus 0.4em\relax
  {IEEE}, 2020, pp. 573--584.

\bibitem{WangYLK21}
R.~Wang, S.~Ye, K.~Li, and S.~Kwong, ``Bayesian network based label correlation
  analysis for multi-label classifier chain,'' \emph{Inf. Sci.}, vol. 554, pp.
  256--275, 2021.

\bibitem{ShanL21}
X.~Shan and K.~Li, ``An improved two-archive evolutionary algorithm for
  constrained multi-objective optimization,'' in \emph{EMO'21: Proc. of the
  11th International Conference on Evolutionary Multicriteria Optimization},
  ser. Lecture Notes in Computer Science, vol. 12654.\hskip 1em plus 0.5em
  minus 0.4em\relax Springer, 2021, pp. 235--247.

\bibitem{LaiL021}
G.~Lai, M.~Liao, and K.~Li, ``Empirical studies on the role of the decision
  maker in interactive evolutionary multi-objective optimization,'' in
  \emph{CEC'21: Proc. of the 2021 IEEE Congress on Evolutionary
  Computation}.\hskip 1em plus 0.5em minus 0.4em\relax {IEEE}, 2021, pp.
  185--192.

\bibitem{LiLLM21}
L.~Li, Q.~Lin, K.~Li, and Z.~Ming, ``Vertical distance-based clonal selection
  mechanism for the multiobjective immune algorithm,'' \emph{Swarm Evol.
  Comput.}, vol.~63, p. 100886, 2021.

\bibitem{WuKJLZ17}
M.~Wu, S.~Kwong, Y.~Jia, K.~Li, and Q.~Zhang, ``Adaptive weights generation for
  decomposition-based multi-objective optimization using gaussian process
  regression,'' in \emph{GECCO'17: Proc. of the 2017 Genetic and Evolutionary
  Computation Conference}.\hskip 1em plus 0.5em minus 0.4em\relax {ACM}, 2017,
  pp. 641--648.

\bibitem{LiCSY19}
K.~Li, R.~Chen, D.~A. Savic, and X.~Yao, ``Interactive decomposition
  multiobjective optimization via progressively learned value functions,''
  \emph{{IEEE} Trans. Fuzzy Syst.}, vol.~27, no.~5, pp. 849--860, 2019.

\bibitem{LiLDMY20}
K.~Li, M.~Liao, K.~Deb, G.~Min, and X.~Yao, ``Does preference always help? {A}
  holistic study on preference-based evolutionary multiobjective optimization
  using reference points,'' \emph{{IEEE} Trans. Evol. Comput.}, vol.~24, no.~6,
  pp. 1078--1096, 2020.

\bibitem{WuLKZ20}
M.~Wu, K.~Li, S.~Kwong, and Q.~Zhang, ``Evolutionary many-objective
  optimization based on adversarial decomposition,'' \emph{{IEEE} Trans.
  Cybern.}, vol.~50, no.~2, pp. 753--764, 2020.

\bibitem{PruvostDLL020}
G.~Pruvost, B.~Derbel, A.~Liefooghe, K.~Li, and Q.~Zhang, ``On the combined
  impact of population size and sub-problem selection in {MOEA/D},'' in
  \emph{EvoCOP'20: Proc. of the 20th European Conference on Evolutionary
  Computation in Combinatorial Optimization}, ser. Lecture Notes in Computer
  Science, vol. 12102.\hskip 1em plus 0.5em minus 0.4em\relax Springer, 2020,
  pp. 131--147.

\bibitem{XuLA22}
J.~Xu, K.~Li, and M.~Abusara, ``Preference based multi-objective reinforcement
  learning for multi-microgrid system optimization problem in smart grid,''
  \emph{Memetic Comput.}, vol.~14, no.~2, pp. 225--235, 2022.

\bibitem{LiLL22}
S.~Li, K.~Li, and W.~Li, ``Do we really need to use constraint violation in
  constrained evolutionary multi-objective optimization?'' in \emph{PPSN'22:
  Proc. of the 17th International Conference on Parallel Problem Solving from
  Nature}, ser. Lecture Notes in Computer Science, vol. 13399.\hskip 1em plus
  0.5em minus 0.4em\relax Springer, 2022, pp. 124--137.

\bibitem{ZhouLM22}
S.~Zhou, K.~Li, and G.~Min, ``Attention-based genetic algorithm for adversarial
  attack in natural language processing,'' in \emph{PPSN'22: Proc. of 17th
  International Conference on Parallel Problem Solving from Nature}, ser.
  Lecture Notes in Computer Science, vol. 13398.\hskip 1em plus 0.5em minus
  0.4em\relax Springer, 2022, pp. 341--355.

\bibitem{ChenLTL22}
L.~Chen, H.~Liu, K.~C. Tan, and K.~Li, ``Transfer learning-based parallel
  evolutionary algorithm framework for bilevel optimization,'' \emph{{IEEE}
  Trans. Evol. Comput.}, vol.~26, no.~1, pp. 115--129, 2022.

\bibitem{Williams0M22}
P.~N. Williams, K.~Li, and G.~Min, ``Black-box adversarial attack via
  overlapped shapes,'' in \emph{{GECCO} '22: Genetic and Evolutionary
  Computation Conference, Companion Volume, Boston, Massachusetts, USA, July 9
  - 13, 2022}.\hskip 1em plus 0.5em minus 0.4em\relax {ACM}, 2022, pp.
  467--468.

\bibitem{FanLT20}
X.~Fan, K.~Li, and K.~C. Tan, ``Surrogate assisted evolutionary algorithm based
  on transfer learning for dynamic expensive multi-objective optimisation
  problems,'' in \emph{CEC'20: Proc. of the 2020 {IEEE} Congress on
  Evolutionary Computation}.\hskip 1em plus 0.5em minus 0.4em\relax {IEEE},
  2020, pp. 1--8.

\bibitem{ZitzlerK04}
E.~Zitzler and S.~K{\"{u}}nzli, ``Indicator-based selection in multiobjective
  search,'' in \emph{{PPSN}'04: Proc. of the 8th International Conference on
  Parallel Problem Solving from Nature}, 2004, pp. 832--842.

\bibitem{ZhangL07}
Q.~Zhang and H.~Li, ``{MOEA/D:} {A} multiobjective evolutionary algorithm based
  on decomposition,'' \emph{{IEEE} Trans. Evol. Comput.}, vol.~11, no.~6, pp.
  712--731, 2007.

\bibitem{ZitzlerT99}
E.~Zitzler and L.~Thiele, ``Multiobjective evolutionary algorithms: a
  comparative case study and the strength pareto approach,'' \emph{{IEEE}
  Trans. Evol. Comput.}, vol.~3, no.~4, pp. 257--271, 1999.

\bibitem{PeiHXLWW20}
J.~Pei, P.~Hong, K.~Xue, D.~Li, D.~S.~L. Wei, and F.~Wu, ``Two-phase virtual
  network function selection and chaining algorithm based on deep learning in
  sdn/nfv-enabled networks,'' \emph{{IEEE} J. Sel. Areas Commun.}, vol.~38,
  no.~6, pp. 1102--1117, 2020.

\bibitem{LiuZDLGZ18}
X.~F. Liu, Z.~Zhan, J.~D. Deng, Y.~Li, T.~Gu, and J.~Zhang, ``An energy
  efficient ant colony system for virtual machine placement in cloud
  computing,'' \emph{{IEEE} Trans. Evol. Comput.}, vol.~22, no.~1, pp.
  113--128, 2018.

\bibitem{CharismiadisTPM20}
A.~Charismiadis, D.~Tsolkas, N.~I. Passas, and L.~F. Merakos, ``A metaheuristic
  approach for minimizing service creation time in slice-enabled networks,'' in
  \emph{{ICC}'20: Proc. of the IEEE 2020 International Conference on
  Communications}.

\bibitem{RuanLDL20}
X.~Ruan, K.~Li, B.~Derbel, and A.~Liefooghe, ``Surrogate assisted evolutionary
  algorithm for medium scale multi-objective optimisation problems,'' in
  \emph{{GECCO} '20: Proc. of 2020 Genetic and Evolutionary Computation
  Conference}.\hskip 1em plus 0.5em minus 0.4em\relax {ACM}, 2020, pp.
  560--568.

\bibitem{SunL20}
L.~Sun and K.~Li, ``Adaptive operator selection based on dynamic thompson
  sampling for {MOEA/D},'' in \emph{PPSN'20: Proc. of 16th International
  Conference on Parallel Problem Solving from Nature}, ser. Lecture Notes in
  Computer Science, vol. 12270.\hskip 1em plus 0.5em minus 0.4em\relax
  Springer, 2020, pp. 271--284.

\bibitem{LiNGY22}
K.~Li, H.~Nie, H.~Gao, and X.~Yao, ``Posterior decision making based on
  decomposition-driven knee point identification,'' \emph{{IEEE} Trans. Evol.
  Comput.}, vol.~26, no.~6, pp. 1409--1423, 2022.

\bibitem{WuLKZZ19}
\BIBentryALTinterwordspacing
M.~Wu, K.~Li, S.~Kwong, Q.~Zhang, and J.~Zhang, ``Learning to decompose: {A}
  paradigm for decomposition-based multiobjective optimization,'' \emph{{IEEE}
  Trans. Evol. Comput.}, vol.~23, no.~3, pp. 376--390, 2019. [Online].
  Available: \url{https://doi.org/10.1109/TEVC.2018.2865931}
\BIBentrySTDinterwordspacing

\bibitem{WuLKZZ17}
M.~Wu, K.~Li, S.~Kwong, Y.~Zhou, and Q.~Zhang, ``Matching-based selection with
  incomplete lists for decomposition multiobjective optimization,''
  \emph{{IEEE} Trans. Evol. Comput.}, vol.~21, no.~4, pp. 554--568, 2017.

\bibitem{LiC23}
K.~Li and R.~Chen, ``Batched data-driven evolutionary multiobjective
  optimization based on manifold interpolation,'' \emph{{IEEE} Trans. Evol.
  Comput.}, vol.~27, no.~1, pp. 126--140, 2023.

\bibitem{WilliamsLM23}
P.~N. Williams, K.~Li, and G.~Min, ``Sparse adversarial attack via bi-objective
  optimization,'' in \emph{EMO'23: Proc. of 12th International Conference on
  Evolutionary Multi-Criterion Optimization}, ser. Lecture Notes in Computer
  Science, vol. 13970.\hskip 1em plus 0.5em minus 0.4em\relax Springer, 2023,
  pp. 118--133.

\bibitem{LyuYWHL23}
B.~Lyu, Y.~Yang, S.~Wen, T.~Huang, and K.~Li, ``Neural architecture search for
  portrait parsing,'' \emph{{IEEE} Trans. Neural Networks Learn. Syst.},
  vol.~34, no.~3, pp. 1112--1121, 2023.

\end{thebibliography}
